\documentclass{article}
\usepackage{graphicx}
\usepackage{amsmath}
\usepackage{amsfonts}
\usepackage{amssymb}
\usepackage{indentfirst}
\usepackage{geometry}
\usepackage{hyperref}
\usepackage{subcaption}
\usepackage{authblk}
\usepackage{bm}
\usepackage[numbers]{natbib}

\begin{document}
\title{Hodge Decomposition for Urban Traffic Flow: Limits on Dense OD Graphs and Advantages on Road Networks - Los Angeles Case Study}
\author[1]{Yifei Sun\thanks{Corresponding author: yifeisun@umich.edu}}
\affil[1]{University of Michigan-Ann Arbor, School of Information}

\date{}
\maketitle

\begin{abstract}
We study Hodge decomposition (HodgeRank) for urban traffic flow on two graph representations: dense origin--destination (OD) graphs and road-segment networks. Reproducing the method of Aoki et al.~\cite{aoki2022urban}, we observe that on dense OD graphs the curl and harmonic components are negligible and the potential closely tracks node divergence, limiting the added value of Hodge potentials. In contrast, on a real road network (UTD19, downtown Los Angeles; 15-minute resolution), potentials differ substantially from divergence and exhibit clear morning/evening reversals consistent with commute patterns. We quantify smoothness and discriminability via local/global variances derived from the graph spectrum, and propose flow-aware embeddings that combine topology, bidirectional volume, and net-flow asymmetry for clustering. Code and preprocessing steps are provided to facilitate reproducibility.
\end{abstract}

\section*{Introduction}
Hodge decomposition is a mathematical framework that decomposes flow‐valued data on graphs (or differential forms on manifolds) into three orthogonal components: a gradient component (from a potential), a curl component (local cyclic inconsistencies), and a harmonic component (global cyclic structure). This decomposition enables interpretable analysis of flows over complex network structures. 

The HodgeRank framework applies this decomposition to pairwise comparison / flow data. Given a graph of items with observed pairwise (or origin–destination) flows, HodgeRank separates them into a global ranking (gradient), local inconsistencies (curl), and global cycles (harmonic). This allows both a quantitative ordering of nodes and a diagnostic of where and how the data deviate from pure gradient structure. See also \cite{jiang2011statistical,Schaub2021HONtutorial} for the Hodge-theoretic view on edge flows and higher-order network structure.

I reproduced the method by Aoki et al. in \cite{aoki2022urban} ,which applies Hodge–Kodaira decomposition / HodgeRank to origin-destination flow data without road-network details, deriving a scalar potential field from OD matrices using the same type of OD data. In contrast to their findings showing meaningful “flow potential landscapes,” I found on this dataset that HodgeRank gives results almost indistinguishable from a simple divergence-based ranking: because the OD network is very dense, the curl and harmonic components are negligible, and no visually interesting or interpretable potential sinks/sources emerge beyond what divergence already shows.

Publicly available road-level real-time traffic datasets are very hard to obtain. Fortunately, I found the UTD19 dataset’s downtown Los Angeles data, which shows traffic at relatively fine time resolution (15-minute intervals), enabling this level of detail. I analyzed it using HodgeRank and confirmed that spatial potential differences are clearly visible, reversing between morning and evening rush hours. Because this network approximates a mesh (its topology differs from a pure OD network), expected patterns appear more distinctly.

Additionally, I developed and evaluated clustering strategies for flow graphs that incorporate: (i) network connectivity, (ii) average bidirectional flow, and (iii) the net flow asymmetry / directionality. We compare how embeddings derived from these combined features perform, in contrast to embeddings based only on topology or only on flow magnitude.

\section*{Datasets}

\paragraph{Longitudinal Employer-Household Dynamics Origin-Destination Employment Statistics (LODES) Data}
LODES data, which is collected by the US Census Bureau, contains home-to-work travel patterns, aiding in mapping origin-destination pairs and uncovering commuting trends. This enables us to apply Hodge decomposition on an origin-destination travel flow graph.

\paragraph{UTD19 Data}
UTD19 data is the largest multi-city traffic dataset publicly available. Collected by ETH Zurich, this dataset provides full-day traffic volume data on road networks at intervals of several minutes for 40 cities around the world. We only present the result of Los Angeles in this dataset, because only LA data contains a complete and large enough connected road network that allows a full and interesting results. This dataset enables us to analyze traffic flow on a road segment network using Hodge decomposition during the day. We admit the scarcity of high quality publicly available data in this area.

\paragraph{Puget Sound Regional Council (PSRC) Data}
The PSRC data includes household characteristics, population distribution, and economic factors essential for predicting trip volumes. These demographic inputs enable machine learning models to account for the impact of these variables on travel patterns and behaviors.

\section*{Methodologies}

\paragraph{Spectral Clustering}

The most common motivation for spectral clustering is to minimize graph cut in the graph partitioning problem. Imagine we want to cut some edges to partition nodes into k clusters, then our goal could be to identify edges representing the weakest connections. I aim to minimize the RatioCut or NormalizedCut of a graph partition. The RatioCut formulation follows the definition in Hagen \& Kahng (1992) \cite{HagenKahng1992}, while the NormalizedCut objective was first proposed by Shi \& Malik \cite{ShiMalik2000}. After relaxing the discrete optimization constraints, spectral clustering algorithms such as the one by Ng, Jordan \& Weiss \cite{NgJordanWeiss2002} solve for the smallest non-zero eigenvectors of the (normalized) Laplacian, then run k-means on rows of the resulting eigenvector matrix. As a tutorial reference for various Laplacian definitions and clustering variants, see von Luxburg (2007) \cite{vonLuxburg2007tutorial}.

In the first setting, we try to minimize RatioCut,
$$RatioCut \left( A_1, A_2,...,A_k \right) = \sum_{i=1}^k \frac {W \left( A_i, \bar{A_i} \right) } {|A_i|}$$

where $A_1, A_2,...,A_k$ are one partition of the nodes in the graph. For one cluster $A_m$, construct a normalized indicator vector ${\textit{\textbf{f}}_{A_m}}_{n \times 1}$. $n$ is the number of nodes in the graph.

$$ \left( \textit{\textbf{f}}_{A_m} \right)_{i} = 
\begin{cases} 
    0, & \text{if } v_i \notin A_m \\
    \frac{1}{\sqrt{|A_m|}}, & \text{if } v_i \in A_m
\end{cases} $$

$${\textit{\textbf{f}}_{A_m}}^T \textit{\textbf{f}}_{A_m} = |A_m|  \frac{1}{\sqrt{|A_m|}}  \frac{1}{\sqrt{|A_m|}} = 1$$

$${\textit{\textbf{f}}_{A_m}}^T \textit{\textbf{f}}_{A_n}  = 0$$

$$\textit{\textbf{F}} = {\textit{\textbf{F}}}_{n \times k} = \begin{bmatrix}
    \textit{\textbf{f}}_{A_1} & \textit{\textbf{f}}_{A_2} & ... & \textit{\textbf{f}}_{A_k} \\
\end{bmatrix}$$

$${{\textit{\textbf{F}}}_{n \times k}}^T {\textit{\textbf{F}}}_{n \times k}={\textit{\textbf{I}}}_{k \times k} = \textit{\textbf{I}}$$

From the property of graph Laplacian, we have 

$${\textit{\textbf{f}}_{A_m}}^T \textit{\textbf{L}} \textit{\textbf{f}}_{A_m} = \frac {1} {2} \sum_{i=1}^n \sum_{j=1}^n w_{ij} \left( \left( \textit{\textbf{f}}_{A_m} \right)_{i} - \left( \textit{\textbf{f}}_{A_m} \right)_{j} \right)^2 = \frac {W \left( A_m, \bar{A_m} \right) } {|A_m|} $$

The RatioCut goal becomes

$$\mathrm{tr}(\textit{\textbf{F}}^T \textit{\textbf{L}} \textit{\textbf{F}}) = \mathrm{tr}(\textit{\textbf{F}}_{n \times k}^T \textit{\textbf{L}}_{n \times n} \textit{\textbf{F}}_{n \times k}) = \sum_{m=1}^k {{\textit{\textbf{f}}_{A_m}}^T \textit{\textbf{L}} \textit{\textbf{f}}_{A_m}} = \sum_{m=1}^k \frac {W \left( A_m, \bar{A_m} \right) } {|A_m|} = RatioCut \left( A_1, A_2,...,A_k \right)$$

The RatioCut optimization problem can be formulated as 

$$ \underset{ \textit{\textbf{F}} \in \mathbb{R}^{n \times k}}{\text{argmin}}  \hspace{0.2cm}  \textnormal{tr}  \left( {{\textit{\textbf{F}}}}^T {\textit{\textbf{L}}} {\textit{\textbf{F}}}  \right) = \textnormal{tr} \left(  {{\textit{\textbf{F}}}_{n \times k}}^T {\textit{\textbf{L}}}_{n \times n} {\textit{\textbf{F}}}_{n \times k}  \right)   \hspace{0.2cm}
 s.t. \textit{\textbf{F}} = {\textit{\textbf{F}}}_{n \times k} = \begin{bmatrix}
    \textit{\textbf{f}}_{A_1} & \textit{\textbf{f}}_{A_2} & ... & \textit{\textbf{f}}_{A_k} \\
\end{bmatrix}   \hspace{0.2cm} \left( \textit{\textbf{f}}_{A_m} \right)_{i} = 
\begin{cases} 
    0, & \text{if } v_i \notin A_m \\
    \frac{1}{\sqrt{|A_m|}}, & \text{if } v_i \in A_m
\end{cases}$$

Here the matrix \textit{\textbf{F}} encodes the partition information of the nodes, which is similar to a one-hot encoding matrix. This matrix should be sparse. Because this problem is by nature a combinatorics problem, it is NP hard and is difficult to get the exact optimum. We relax the constraint a bit and reformulate the problem as the following. The result of the reformulated problem has solution approximate to the original optimum. This result could be quite different from the real optimum in graph cut problem.

$$ \underset{ \textit{\textbf{F}} \in \mathbb{R}^{n \times k}}{\text{argmin}}  \hspace{0.2cm} \textnormal{tr} { \left( {{\textit{\textbf{F}}}}^T {\textit{\textbf{L}}} {\textit{\textbf{F}}} \right) } = \textnormal{tr} \left( {{{\textit{\textbf{F}}}_{n \times k}}^T {\textit{\textbf{L}}}_{n \times n} {\textit{\textbf{F}}}_{n \times k}} \right)   \hspace{0.5cm}
 s.t. {{\textit{\textbf{F}}}_{n \times k}}^T {\textit{\textbf{F}}}_{n \times k}={\textit{\textbf{I}}}_{k \times k} = \textit{\textbf{I}}$$

Here the problem becomes somewhat similar to a Rayleigh quotient problem, we just need to find a series of $k$ orthonormal bases that minimize ${{\textit{\textbf{F}}}}^T {\textit{\textbf{L}}} {\textit{\textbf{F}}}$. We could just use the $k$ eigenvectors corresponding to the $k$ smallest non-zero eigenvalues of \textit{\textbf{L}}. The graph Laplacian  \textit{\textbf{L}} has at least one eigenvalue 0 and its associated eigenvector is all 1 vector, which is useless in partitioning the graph.  Note that the matrix \textit{\textbf{F}} we get from eigenvectors of \textit{\textbf{L}} are neither similar to one-hot encoding nor sparse anymore.

In the second setting, we try to minimize NormalizedCut,
$$NormalizedCut \left( A_1, A_2,...,A_k \right) = \sum_{i=1}^k \frac {W \left( A_i, \bar{A_i} \right) } {vol(A_i)}$$

The optimization problem becomes

$$ \underset{ \textit{\textbf{F}} \in \mathbb{R}^{n \times k}}{\text{argmin}}   \hspace{0.2cm} \textnormal{tr} { \left(   {{\textit{\textbf{F}}}}^T {\textit{\textbf{L}}} {\textit{\textbf{F}}}  \right) } =  \textnormal{tr} { \left( {{\textit{\textbf{F}}}_{n \times k}}^T {\textit{\textbf{L}}}_{n \times n} {\textit{\textbf{F}}}_{n \times k}  \right) } \hspace{0.2cm}
 s.t. \textit{\textbf{F}} = {\textit{\textbf{F}}}_{n \times k} = \begin{bmatrix}
    \textit{\textbf{f}}_{A_1} & \textit{\textbf{f}}_{A_2} & ... & \textit{\textbf{f}}_{A_k} \\
\end{bmatrix}   \hspace{0.2cm} \left( \textit{\textbf{f}}_{A_m} \right)_{i} = 
\begin{cases} 
    0, & \text{if } v_i \notin A_m \\
    \frac{1}{\sqrt{vol(A_i)}}, & \text{if } v_i \in A_m
\end{cases}$$

By relaxing the constraints, we have

$$ \underset{ \textit{\textbf{F}} \in \mathbb{R}^{n \times k}}{\text{argmin}}  \hspace{0.2cm} \textnormal{tr} { \left(  {{\textit{\textbf{F}}}}^T {\textit{\textbf{L}}} {\textit{\textbf{F}}} = {{\textit{\textbf{F}}}_{n \times k}}^T {\textit{\textbf{L}}}_{n \times n} {\textit{\textbf{F}}}_{n \times k}  \right) }  \hspace{0.5cm}
 s.t. {{\textit{\textbf{F}}}_{n \times k}}^T {\textit{\textbf{D}}}_{n \times n} {\textit{\textbf{F}}}_{n \times k}={\textit{\textbf{I}}}_{k \times k} = \textit{\textbf{I}}$$

This is similar to a generalized Rayleigh quotient problem, we let $\tilde{\textit{\textbf{F}}} = {\textit{\textbf{D}}}^{\frac {1} {2}} \textit{\textbf{F}}$

The optimization problem becomes

$$ \underset{ \tilde{\textit{\textbf{F}}} \in \mathbb{R}^{n \times k}}{\text{argmin}}  \hspace{0.2cm}   \textnormal{tr} { \left(  \tilde{{{\textit{\textbf{F}}}}}^T {\textit{\textbf{D}}}^{-\frac {1} {2}} {\textit{\textbf{L}}} {\textit{\textbf{D}}}^{-\frac {1} {2}} \tilde{{\textit{\textbf{F}}}}  \right) } \hspace{0.5cm}
 s.t. \tilde{{{\textit{\textbf{F}}}}}^T \tilde{{\textit{\textbf{F}}}}= \textit{\textbf{I}}$$

 Here we just take the $k$ eigenvectors associated to the smallest non-zero eigenvalues of the normalized graph Laplacian ${\textit{\textbf{D}}}^{-\frac {1} {2}} {\textit{\textbf{L}}} {\textit{\textbf{D}}}^{-\frac {1} {2}}$ as columns of $\tilde{\textit{\textbf{F}}}$ and we have $\textit{\textbf{F}} = {\textit{\textbf{D}}}^{-\frac {1} {2}} \tilde{\textit{\textbf{F}}}$.

 For spectral clustering, there is a further step, which is to use $\textit{\textbf{F}}$ as embedding in Euclidean space for each node and run K-means clustering. This embedding involving the eigenvectors of graph Laplacian is called Laplacian eigenmap, introduced by Belkin \& Niyogi \cite{BelkinNiyogi2003}. This also aligns with the normalization and relaxation steps in Ng, Jordan \& Weiss \cite{NgJordanWeiss2002}. The similarity matrix or adjacency matrix of the graph should be constructed wisely to reflect the similarities between nodes.

 Another explanation of the spectral clustering method is in topology. The dimension of the kernel of the graph Laplacian, which is the multiplicity of eigenvalue 0, is the number of 0-holes in the graph (the number of connected components). For a well-connected graph, its non-zero eigenvalues are far from 0, while graphs with several obvious clusters will have some eigenvalues very close to 0. Eigenvectors associated with eigenvalues close to 0 indicate the almost components in the graph, and are a good way of identifying clusters in a graph. Spectral properties of the Laplacian matrix (e.g. algebraic connectivity given by the smallest non-zero eigenvalue—aka the Fiedler value) play an important role in determining partition quality and smoothness of potentials on graphs. The foundational results on this topic go back to Fiedler (1973) \cite{Fiedler1973}, and are treated in depth in Chung’s monograph on spectral graph theory \cite{Chung1997}.

For LODES origin-destination data, we could construct an adjacency matrix $\textit{\textbf{A}}$ whose entries are travel volume between a directed origin-destination pair. Because spectral clustering requires the adjacency matrix to be symmetric, we use the mean of volumes on the two opposite origin-destination pairs as the connectedness strength between the two nodes, i.e. 

$$\tilde{\textit{\textbf{A}}} = \frac {\textit{\textbf{A}}+{\textit{\textbf{A}}}^T} {2}$$

We then used the Gaussian kernel to transform the average volumes on edges to pairwise similarities. Larger volume means larger similarity.

This spectral clustering of nodes on the graph does not take the direction and asymmetry of the flows into account, and only uses the average flow volume as the strength of connection between two nodes.

\paragraph{Hodge Decomposition on Graphs}

\begin{figure}[htbp]
    \centering
    \includegraphics[width=0.8\textwidth]{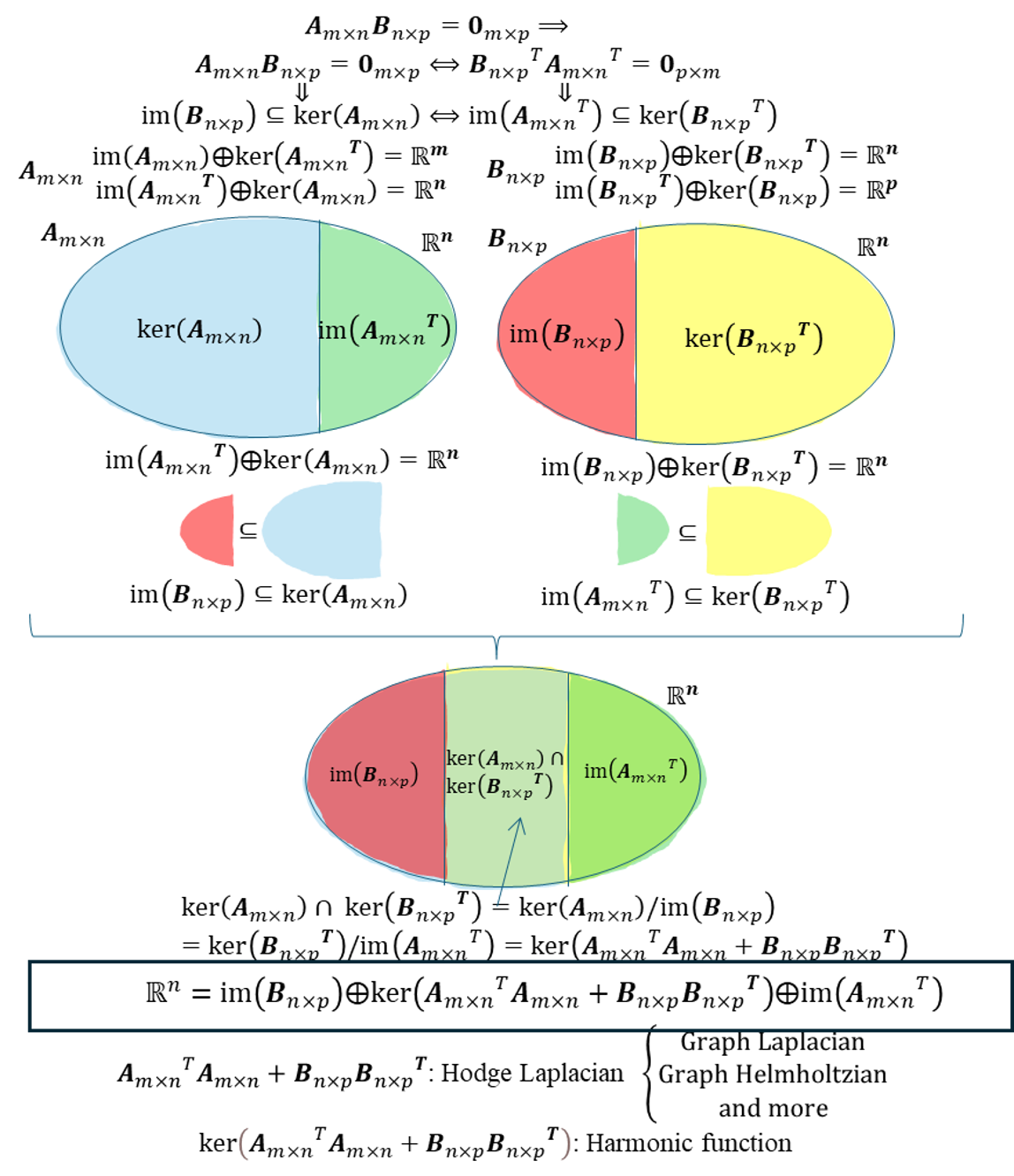}  
    \caption{Illustration of the Derivation of Hodge Decomposition on Graphs (Three Mutually Orthogonal Subspaces Whose Direct Sum is the Intermediate Vector Space)}
    \label{fig:vertical}
\end{figure}

From a pure linear algebra perspective, we could have an intuitive derivation of Hodge decomposition on graphs. Figure \ref{fig:vertical} illustrates such derivation. This derivation is based on the basic facts of the four fundamental subspaces of linear transformation. The only assumption we use here is the composition of two linear transformations $\textit{\textbf{A}}_{m \times n}$, $\textit{\textbf{B}}_{n \times p}$ always gives a zero vector, i.e. $\textit{\textbf{A}}_{m \times n}\textit{\textbf{B}}_{n \times p}=\textbf{0}_{m \times p}$. In topology, this composition that gives zero as its result implies that the boundary of a boundary is always empty. This assumption solely gives the result that the intermediate vector spaces (of dimension $n$) can be decomposed into three mutually orthogonal subspaces, i.e. $\mathrm{im} \left( \textit{\textbf{B}}_{n \times p} \right)$, $\mathrm{ker} \left( \textit{\textbf{A}}_{m \times n}^T \textit{\textbf{A}}_{m \times n} + \textit{\textbf{B}}_{n \times p} \textit{\textbf{B}}_{n \times p}^T \right)$, $\mathrm{im} \left( \textit{\textbf{A}}_{m \times n}^T \right)$,  and when matrices $\textit{\textbf{A}}_{m \times n}$, $\textit{\textbf{B}}_{n \times p}$ are assigned to be certain meaningful matrices in the algebraic context, this decomposition becomes Hodge decomposition on graphs. For a bridge from continuous Hodge theory to discrete operators used on graphs, see DEC \cite{DesbrunHiraniLeokMarsden2005DEC} and FEEC \cite{ArnoldFalkWinther2006FEEC}.

Next we need several more definitions of the operators that are meaningful in the context of discrete vector field (calculus on graphs), i.e. $\mathrm{grad}$, $\mathrm{curl}$, $\mathrm{div}$, and $\mathrm{curl}^*$. 

$$\textit{\textbf{g}} \in \mathbb{R}^{|\textit{E}|},  
 \textit{\textbf{f}} \in \mathbb{R}^{|\textit{V}|},  
 \mathrm{grad} \in \mathbb{R}^{|\textit{E}| \times |\textit{V}|},  \hspace{0.5cm} 
 \mathrm{grad} \textit{\textbf{f}}_{|\textit{V}| \times 1} = \textit{\textbf{g}}_{|\textit{E}| \times 1},  \hspace{0.5cm} \textit{\textbf{g}}_{e_{i,j}} = {\left( \mathrm{grad} \textit{\textbf{f}} \right)}_{e_{i,j}} = f_{v_j} - f_{v_i}$$

$$\textit{\textbf{g}} \in \mathbb{R}^{|\textit{E}|},  
 \textit{\textbf{f}} \in \mathbb{R}^{|\textit{V}|},  
 \mathrm{div} \in \mathbb{R}^{|\textit{V}| \times |\textit{E}|},  \hspace{0.5cm} 
 \mathrm{div} \textit{\textbf{g}}_{|\textit{E}| \times 1} = \textit{\textbf{f}}_{|\textit{V}| \times 1},  \hspace{0.5cm} \textit{\textbf{f}}_{v_{i}} = {\left( \mathrm{div} \textit{\textbf{g}} \right)}_{v_{i}} = \sum_{j} g_{e_{i,j}}$$

$$\textit{\textbf{h}} \in \mathbb{R}^{|\textit{T}|},  
 \textit{\textbf{g}} \in \mathbb{R}^{|\textit{E}|},  
 \mathrm{curl} \in \mathbb{R}^{|\textit{T}| \times |\textit{E}|},  \hspace{0.5cm} 
 \mathrm{curl} \textit{\textbf{g}}_{|\textit{E}| \times 1} = \textit{\textbf{h}}_{|\textit{T}| \times 1},  \hspace{0.5cm} \textit{\textbf{h}}_{t_{i,j,k}} = {\left( \mathrm{curl} \textit{\textbf{g}} \right)}_{t_{i,j,k}} = g_{e_{i,j}} + g_{e_{j,k}} + g_{e_{k,i}} = g_{e_{i,j}} - g_{e_{i,k}} + g_{e_{j,k}}$$

$$\mathrm{-div} = \mathrm{grad}^*$$

$$\mathrm{grad} \circ \mathrm{curl} = 0,  \hspace{0.5cm} \mathrm{curl}^* \circ \mathrm{-div} = 0$$

These four operators in this discrete context are in matrix form. They are some low-dimension special cases of boundary operators and coboundary operators in algebraic topology. Hodge decomposition is the decomposition of cochains defined on any dimensional simplicial complex (functions defined on k-cliques). The boundary relationship between k-simplicial complex and k+1-simplicial complex is described by the boundary/coboundary operator. $\mathrm{grad}$ and $\mathrm{-div}$ are the adjoints (Hermitian conjugates) of each other, and are the lowest order boundary operators. They are merely the incidence matrix and its transpose in graph theory. The composition of $\mathrm{grad}$, and $\mathrm{curl}$ and composition of $\mathrm{-div}$, and $\mathrm{curl}^*$ always give the result of zero, which can be easily verified. This unveils the more fundamental result that the boundary of a boundary is zero.

In general, cochains are functions defined on simplices $\textit{X} \rightarrow \mathbb{R}$, for k-simplicial complex, they can be represented as a vector, i.e. $\mathbb{R}^{|\textit{V}|}$, $\mathbb{R}^{|\textit{E}|}$, $\mathbb{R}^{|\textit{T}|}$, etc. The sign of the value of the function is related to the parity of the permutation.
$$\textit{f} \left(  \left[  i_{p(0)},...,i_{p(k)} \right]  \right) = \mathrm{sign}(p) \textit{f} \left(  \left[  i_0,...,i_k \right]  \right) $$

$ i_{p(0)},...,i_{p(k)} $ is a permutation of $i_0,...,i_k$. If it is an odd permutation, $\mathrm{sign}(p)=-1$; if it is a even permutation, $\mathrm{sign}(p)=1$

Coboundary operator $ \delta_k $ is the linear map from lower order cochains to higher order cochains and is defined as 

$$\left( \delta_k \textit{f} \right) \left( i_0,i_1,...,i_{k+1} \right) := \sum_{j=0}^{k+1} {(-1)}^j \textit{f} \left( i_0,...,i_{j-1},i_{j+1},...,i_{k+1} \right) $$

We can see the above defined operators are merely special cases of the definitions here.

From the above derivation we have 

$$\textit{\textbf{A}}_{m \times n}\textit{\textbf{B}}_{n \times p}=\textbf{0}_{m \times p} \longrightarrow \mathbb{R}^{n} = \mathrm{im} \left( \textit{\textbf{B}}_{n \times p} \right) \oplus \mathrm{ker} \left( \textit{\textbf{A}}_{m \times n}^T \textit{\textbf{A}}_{m \times n} + \textit{\textbf{B}}_{n \times p} \textit{\textbf{B}}_{n \times p}^T \right) \oplus \mathrm{im} \left( \textit{\textbf{A}}_{m \times n}^T \right)$$

By limiting the context to node space, edge space and triangle space, we could replace $\textit{\textbf{A}}_{m \times n}$, $\textit{\textbf{B}}_{n \times p}$, $\textit{\textbf{A}}_{m \times n}^T$, and $\textit{\textbf{B}}_{n \times p}^T$ with $\mathrm{curl}$, $\mathrm{grad}$, $\mathrm{curl}^*$, and $\mathrm{-div}$ and then get 

$$\mathrm{curl}  \hspace{0.2cm}  \mathrm{grad}=\textbf{0}_{m \times p} \longrightarrow \mathbb{R}^{|\textit{E}|} = \mathrm{im} \left( \mathrm{grad} \right) \oplus \mathrm{ker} \left( \mathrm{curl}^* \circ \mathrm{curl} + \mathrm{grad} \circ \left( \mathrm{-div}\right) \right) \oplus \mathrm{im} \left( \mathrm{curl}^* \right)$$
$$= \mathrm{im} \left( \mathrm{grad} \right) \oplus \mathrm{ker} \left(-\mathrm{grad} \circ \mathrm{div} + \mathrm{curl}^* \circ \mathrm{curl} \right) \oplus \mathrm{im} \left( \mathrm{curl}^* \right)$$

$$ \mathrm{graph  \hspace{0.2cm} Laplacian} : -\mathrm{div} \circ \mathrm{grad}$$

$$ \mathrm{graph  \hspace{0.2cm} Helmholtzian} : -\mathrm{grad} \circ \mathrm{div} + \mathrm{curl}^* \circ \mathrm{curl}$$

$\mathrm{ker} \left(-\mathrm{grad} \circ \mathrm{div} + \mathrm{curl}^* \circ \mathrm{curl} \right)$ is called the 1st homology group, which are cycles (which have empty boundary) but not the boundary of anything else. In other words, they are cycles longer than 3. (Cycles of length 3 would be the boundary of an triangle.)
$ \mathrm{dim} \left( \mathrm{ker} \left(-\mathrm{grad} \circ \mathrm{div} + \mathrm{curl}^* \circ \mathrm{curl} \right) \right)$ is the number of order 2 holes in the graph, also called Betti number.

\begin{figure}[htbp]
    \centering
    \includegraphics[width=0.8\textwidth]{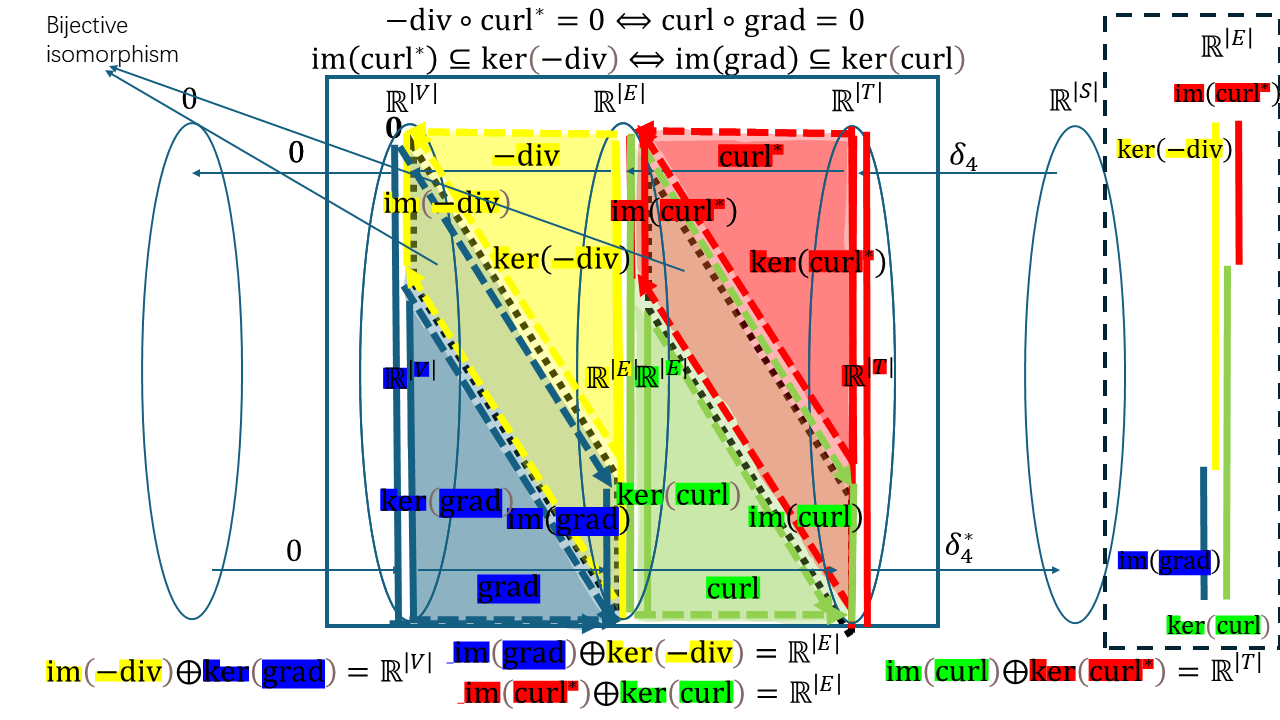}  
    \caption{Illustration of Hodge Decomposition}
    \label{fig:Hodge decomposition illustration}
\end{figure}

\begin{figure}[htbp]
    \centering
    \includegraphics[width=0.8\textwidth]{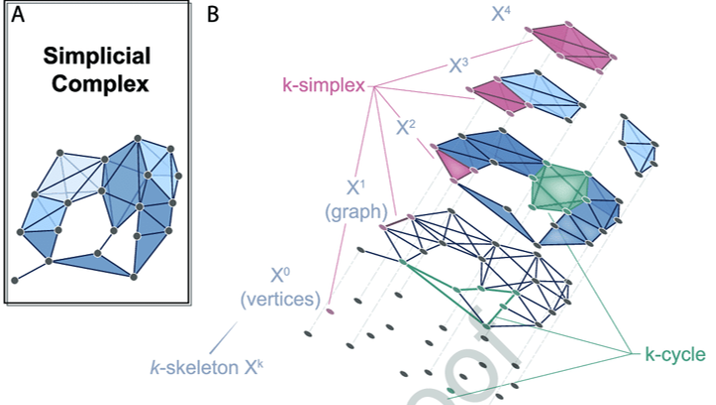}  
    \caption{Illustration of Simplex and Simplicial Complex in Graphs}
    \label{fig:simplicial complex}
\end{figure}

Figure \ref{fig:Hodge decomposition illustration} is an illustration of the Hodge decomposition we used in the project. It is a discrete analogy of the Hodge decomposition of a continuous vector field. Here we define three spaces on the graph, which are node space, edge space, and triangle space. In algebraic topology, nodes, edges, and triangles (1-clique, 2-clique, and 3-clique in graph theory) correspond to 0-simplicial complex, 1-simplicial complex, and 2-simplicial complex. Every node, edge, and triangle (0-simplex, 1-simplex, 2-simplex) can have a value defined on it, and thus create three vector spaces associated with nodes, edges, and triangles respectively. The relation between cliques in graphs and simplices in topology is shown in Figure \ref{fig:simplicial complex}. Boundary operators and their adjoints define the relationship between those three spaces.

These three components have respective meanings, i.e.

$$ \mathrm{Edge  \hspace{0.2cm} Flow = \lefteqn{\overbrace{\phantom{conservative \oplus  \mathrm{solenoidal  \hspace{0.2cm} and  \hspace{0.2cm} irrotational}}}^{\mathrm{irrotational}}}conservative \oplus \underbrace{\mathrm{solenoidal  \hspace{0.2cm} and  \hspace{0.2cm} irrotational} \oplus \mathrm{vorticity}}_{\mathrm{solenoidal }}}$$

$$ \mathrm{Edge  \hspace{0.2cm} Flow = \lefteqn{\overbrace{\phantom{curl-free \hspace{0.1cm} but \hspace{0.1cm} not \hspace{0.1cm} divergence-free \oplus  \mathrm{divergence-free  \hspace{0.1cm} and  \hspace{0.1cm} curl-free}}}^{\mathrm{curl-free}}} curl-free \hspace{0.1cm} but \hspace{0.1cm} not \hspace{0.1cm} divergence-free \oplus \underbrace{\mathrm{divergence-free  \hspace{0.1cm} and  \hspace{0.1cm} curl-free} \oplus \mathrm{divergence-free \hspace{0.1cm} but \hspace{0.1cm} not \hspace{0.1cm} curl-free}}_{\mathrm{divergence-free}}}$$

In HodgeRank, only the $\mathrm{im} \left( \mathrm{grad}\right)$ (conservative) component is used to calculate the ranking values of nodes. For $\mathrm{im} \left( \mathrm{grad}\right)$ component of the edge flow, the flow value on every edge (predefined with an orientation) is the difference between the values of the two end nodes. The values defined on nodes that create this edge flow component is called the potential of this graph. In this sense, $\mathrm{im} \left( \mathrm{grad}\right)$ is perfectly consistent for every pair of neighboring nodes. $\mathrm{im} \left( \mathrm{curl}^* \right)$ is used as the measurement for local inconsistency (inconsistency between three mutually connected nodes). $\mathrm{ker} \left(-\mathrm{grad} \circ \mathrm{div} + \mathrm{curl}^* \circ \mathrm{curl} \right)$ is used as measurement for global inconsistency (inconsistency on cycles longer than 3). Larger $\mathrm{im} \left( \mathrm{grad}\right)$ component in the edge flow indicates the high rankability of the graph, and vice versa. Curl measures local/global inconsistencies as in HodgeRank \citep{jiang2011statistical} and in tutorials on Hodge Laplacians for higher-order networks \citep{lim2020hodge,Schaub2021HONtutorial}.

\begin{figure}[htbp]
    \centering
    \includegraphics[width=0.8\textwidth]{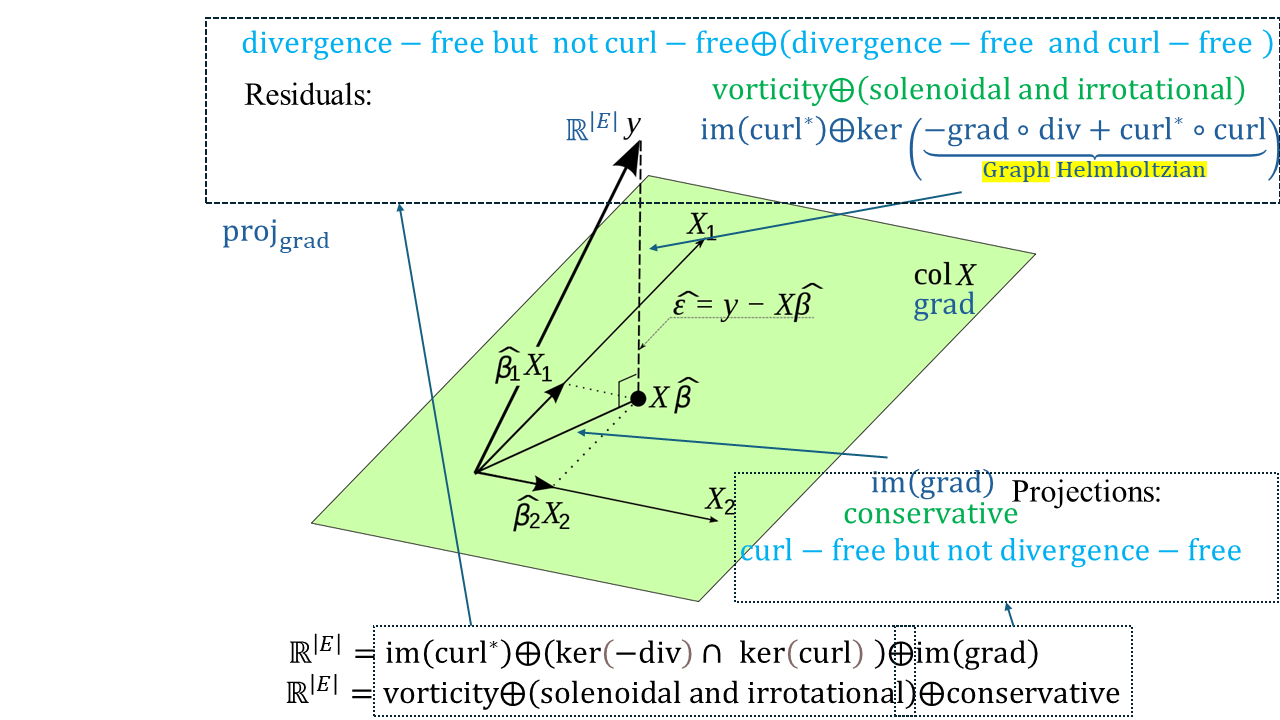}  
    \caption{Illustration of HodgeRank Using Projection}
    \label{fig:Hodge decomposition 2}
\end{figure}

To get the potentials on the nodes of a graph given the edge flows, we would just project the edge flows onto the node space in order to get the potential vector that could create the closest purely gradient edge flows to the raw edge flows (because of the orthogonality condition). This is similar to solving linear regression using least squares, in which $\textit{\textbf{y}}$, $\textit{\textbf{X}}$, and $\textit{\textbf{$\beta$}}$ in $\textit{\textbf{y}} = \textit{\textbf{X}} \textit{\textbf{$\beta$}} + \textit{\textbf{$\epsilon$}}$ are replaced by $\textit{\textbf{f}}$ (edge flows defined on edges), $\mathrm{grad}$ (gradient operator), and $\textit{\textbf{r}}$ (potentials defined on nodes) as illustrated in Figure \ref{fig:Hodge decomposition 2}. The difference is that in linear regression we assume independent columns in the matrix $\textit{\textbf{X}}$, and we try to avoid multicollinearity. However, this incidence matrix $\mathrm{grad}$ is always neither full column rank nor full row rank, so there are infinite approximate solutions by projection, and these solutions are addition invariant, which makes sense because potential is only meaningful when there is a reference point. Another feature of the matrix $\mathrm{grad}$ is that it is very sparse, therefore we used $\mathrm{scipy.sparse\_ matrix}$ to accelerate computation. Also, this matrix could have a lot more edges than nodes if we assume the graph to be similar to an Erd\H{o}s-Renyi random graph, because the number of edges is quadratic to the number of nodes in this setting, meaning that this matrix is super tall and slim.

To solve this we would normally employ the Moore-Penrose pseudo-inverse of matrix $\mathrm{grad}$ (here denoted as $\partial_1^T$, meaning the adjoint of boundary operator of order 1). Note in HodgeRank, there could be a weight assigned to each edge, and the linear regression is solved using weighted least squares.

$$\min_{\substack{\textit{\textbf{r}} \in \mathbb{R}^{|\textit{V}|}}} {\lVert \textit{\textbf{f}} - \partial_1^T \textit{\textbf{r}} \rVert}_{\textit{\textbf{W}}}^2 $$

The solutions are

$$\textit{\textbf{r}} = {\left( \partial_1 \textit{\textbf{W}}^{\frac {1}{2}} \right)^T}^\dagger \textit{\textbf{W}}^{\frac {1}{2}} \textit{\textbf{f}} + \left( \textit{\textbf{I}} - {\left( \partial_1 \textit{\textbf{W}}^{\frac {1}{2}} \right)^T}^\dagger \left( \partial_1 \textit{\textbf{W}}^{\frac {1}{2}} \right)^T \right) \textit{\textbf{w}} $$

For UTD19 data, we have the traffic volumes in both directions of a road segment. The first step we need to do is to construct the adjacency matrix $\textit{\textbf{A}}$, whose entries are the traffic volume from the start node to the end node of a road segment. We will define the direction of each road segment and get the difference of volumes in both directions, i.e.

$$\tilde{\textit{\textbf{A}}} = \textit{\textbf{A}}-{\textit{\textbf{A}}}^T$$

Note that here an edge of net flow 0 does not have the same effect on the network as that edge does not exist. The nonexistence of an edge between two nodes indicates there is no possible information on the difference between the potentials of the two nodes, while an edge with the same volumes on both directions indicates their potentials are somewhat the same. The structure of the network and the flow values on the network are two different inputs of the model.

Our motivation for using HodgeRank on this traffic flow network is to model it like an electric circuit. People travel in the network from nodes to nodes like electrons travel in a circuit from nodes to nodes. There is a potential distributed across the nodes in the network that drives people to commute from suburbs to downtown during morning rush hour and reversely during evening rush hour. Here the inherent assumption is that the resistance on every road is the same value 1, so the current and voltage on a road segment are the same. 

The effect of HodgeRank method is that when given all edge flows, we could get the potentials on nodes and how confident the potentials are. It achieves an embedding of nodes using edge data.

\section*{Data Preprocessing}
I used ArcGIS to perform data preprocessing on the raw csv dataset from UTD19. The raw data in UTD19 contain the latitude and longitude of every measured point. The measured points are usually at the intersection of roads. A road segment is between two intersections and is represented by four points that include the end and start of that road segment. This means that usually every intersection is represented by many points (the same as the number of road segments connecting this intersection), closely scattered around that intersection. And important task here is to merge all points at one intersection into one node representing the intersection. The ArcGIS tool used here is XY Table-to-Point, Create Buffer, Dissolve Boundaries, and Spatial Join. 

\section*{Results}

\subsection*{Spectral Clustering}

We use the the average of bidirectional travel volumes between two nodes in the PSRC as connection strength on that edge and cluster the graph to 11 clusters. Because the clusters in the result become very disconnected, we add the distance data into the similarity matrix to assign closer nodes with higher similarity. The result is shown in Figure \ref{fig:simplicial complex}. From the results in Appendix B, we can see that increasing weight of the distance factor in the similarity matrix will reduce the disconnectedness in the clusters. The inherent paradox here is that geographical proximity and traveling volume could be mutually exclusive. People may tend to travel to places that are far from their origins. There could be further analysis on traveling volume and travel distance distribution. The criteria for similarity between two nodes in a flow graph could be further explored, for example, average traveling volume, asymmetry in the traveling volume in terms of direction, and geographical proximity. That is part of the reason why we employed HodgeRank in this study.

\begin{figure}[htbp]
    \centering
    \includegraphics[width=0.5\textwidth]{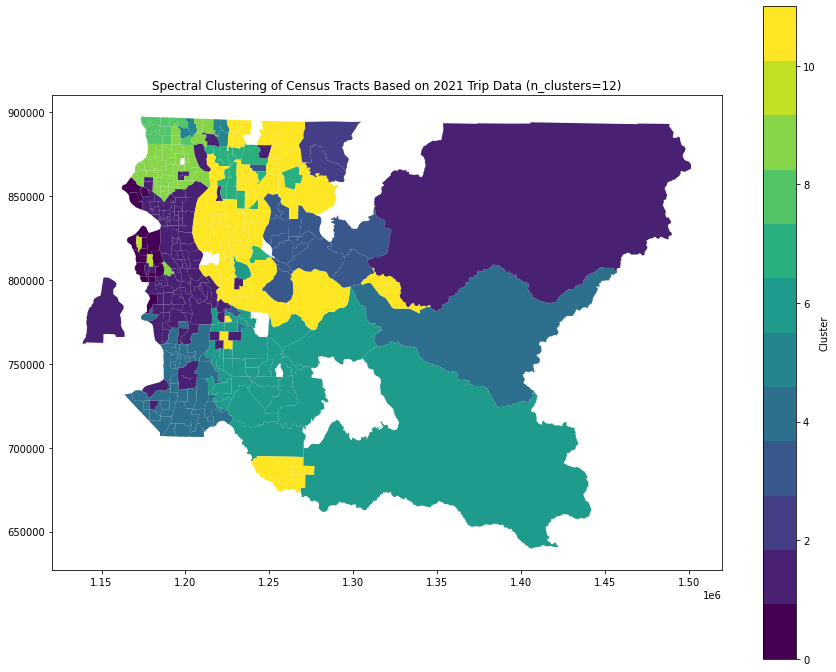}  
    \caption{Spectral Clustering of Census Tracts in King County, Washington Based on PSRC Data }
    \label{fig:cluster}
\end{figure}

\subsection*{HodgeRank Negative Potential and Divergence}

\begin{figure}[htbp]
    \centering
    \includegraphics[width=0.5\textwidth]{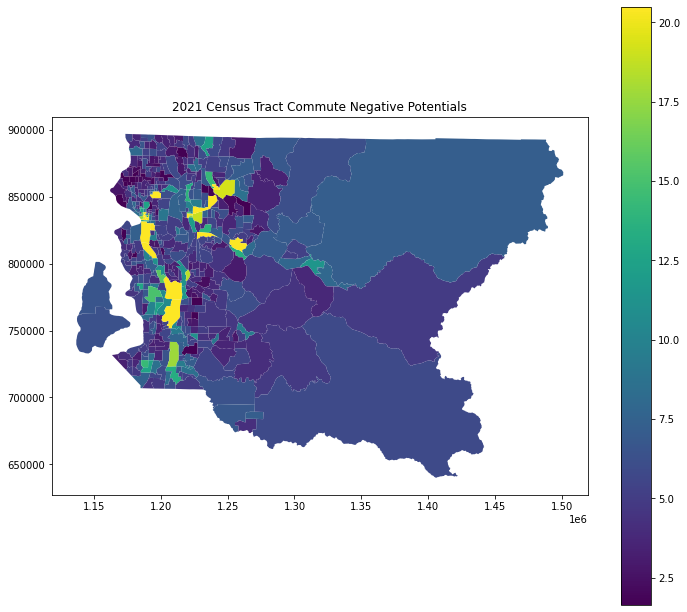}  
    \caption{Negative Potential Values From HodgeRank of King County, Washington Using Work-Home Data From LODES Data}
    \label{fig:hodge}
\end{figure}

The negative potential in King County, Washington from HodgeRank using LODES data is shown in Figure \ref{fig:hodge}. Highlighted areas of high negative potential is where people tend to travel for work, and darker areas indicate residential area. 

In this Hodge decomposition framework, two functions can be defined on nodes, which are potential and divergence. Potential indicates the high or low position of nodes in this flow network. Divergence is simply the net inflow minus outflow in the nodes, indicating the sheer volume into or out of nodes. They are related through the following equations

$$\textit{\textbf{f}}_{|\textit{E}| \times 1} = \textit{\textbf{M}}_{|\textit{E}| \times |\textit{V}|} \textit{\textbf{p}}_{|\textit{V}| \times 1}$$

$$\textit{\textbf{d}}_{|\textit{V}| \times 1} = \textit{\textbf{M}}_{|\textit{E}| \times |\textit{V}|}^T \textit{\textbf{f}}_{|\textit{E}| \times 1}$$

$$\textit{\textbf{d}}_{|\textit{V}| \times 1} = \textit{\textbf{M}}_{|\textit{E}| \times |\textit{V}|}^T \textit{\textbf{M}}_{|\textit{E}| \times |\textit{V}|} \textit{\textbf{p}}_{|\textit{V}| \times 1} = \textit{\textbf{L}}_{|\textit{V}| \times |\textit{V}|} \textit{\textbf{p}}_{|\textit{V}| \times 1}$$

$$\textit{\textbf{p}}_{|\textit{V}| \times 1} = \textit{\textbf{L}}_{|\textit{V}| \times |\textit{V}|}^{\dagger} \textit{\textbf{d}}_{|\textit{V}| \times 1}$$

\begin{figure}[htbp]
    \centering
    \includegraphics[width=0.5\textwidth]{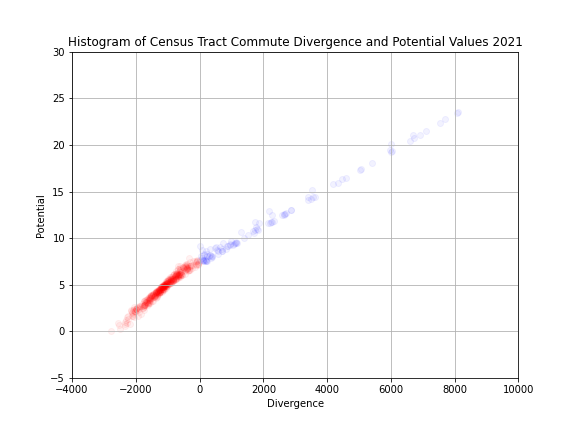}  
    \caption{Negative Potential Values and Divergence of King County, Washington Using Work-Home Data From LODES Data}
    \label{fig:king pot div}
\end{figure}

$\textit{\textbf{f}}_{|\textit{E}| \times 1}$ is gradient flows on edges, $\textit{\textbf{p}}_{|\textit{V}| \times 1}$ is potentials on nodes, $\textit{\textbf{d}}_{|\textit{V}| \times 1}$ is divergences on nodes. $\textit{\textbf{M}}_{|\textit{E}| \times |\textit{V}|}$ is the directed incidence matrix. When there is at most one edge between every pair of nodes on the directed graph, $\textit{\textbf{M}}_{|\textit{E}| \times |\textit{V}|}^T \textit{\textbf{M}}_{|\textit{E}| \times |\textit{V}|}$ is equal to the graph Laplacian $\textit{\textbf{L}}_{|\textit{V}| \times |\textit{V}|}$. In the LODES dataset, we noticed potential and divergence of nodes tend to have a piecewise linear relationship with 0 as its inflection point, as shown in Figure \ref{fig:king pot div}. 

The reason for this piecewise linear relationship is that this network constructed from origin-destination data has some particular features. Nodes in the downtown area have very large degrees and receive large influx during work time, while nodes in suburbs have small degrees and witness net outflow during work time. The graph Laplacian of this network is almost diagonal with diagonal entries at the magnitude of several hundred and non-diagonal entries very sparse and being 0 or -1 (see Appendix C). The diagonal entries of the pseudo inverse of the graph Laplacian are very close to the inverses of diagonal entries of the graph Laplacian. Therefore the slopes for divergence $>$ 0 and divergence $<$ 0 are different. Also, this high linearity indicates that using the potential to rank nodes in the network becomes almost meaningless, as it roughly gives the same ranking as using divergence, and divergence is easier to compute as it only involves doing matrix multiplication while potential requires matrix inversion. This could be a complementary conclusion for   \cite{aoki2022urban} as we used the same dataset and conducted the same tasks.

Because this network only contains origin-destination information, nodes far away could be connected, resulting in high degrees for downtown nodes. This network also ignores the details of trips as people can not teleport from origins to destinations. Therefore, we used a new dataset, UTD19, to apply HodgeRank on a real road network in downtown Los Angeles. Because of the mesh-like shape of the graph, this graph has much fewer edges than the origin-destination network. The results are shown in Appendix D.

We observe that during morning and evening rush hours, the high potential areas and lower potential areas are reversed. As people move from high-potential areas to lower-potential areas, these plots indicate the commute corridor of downtown LA is from downtown to northeast and southwest. From the scatter plots, we notice that, unlike the origin-destination network, potential and divergence here are slightly positively correlated but the significance is much weaker. This is because the nodes in the network have degrees of at most 5, and the non-diagonal entries in the pseudo inverse of the graph Laplacian have values of large magnitudes. And there is not a significant difference in degrees between different nodes. From the histogram, we notice that potential and divergence are of roughly the same magnitude range, but potential distribution is more spread out while divergence distribution is more centered around 0. The result indicates that the values of potential change smoothly across neighboring nodes in the graph while the values of divergence change abruptly across neighboring nodes. To further investigate this property, we develop the concepts of local variance and global variance on graphs.

The sample variance for a sequence of values is 

$$ \frac {1} {n-1} \sum_{i=1}^n {(x_i -\bar{x})}^2 $$

When expressed in vector form,

$$ \frac {1} {n-1} \textit{\textbf{x}}^T \left( \textit{\textbf{I}} - \frac {1} {n} {\textbf{1}}_{n \times n} \right) \textit{\textbf{x}} $$

$n \left( \textit{\textbf{I}} - \frac {1} {n} {\textbf{1}}_{n \times n} \right) = n \textit{\textbf{I}} - {\textbf{1}}_{n \times n}$ can be interpreted as the graph Laplacian of a complete graph of $n$ nodes.

The quadratic form of a graph Laplacian can describe the sum of squared differences of neighboring nodes' values

$${\textit{\textbf{x}}}^T \textit{\textbf{L}} \textit{\textbf{x}} = \frac {1} {2} \sum_{i \sim j} \left( x_{i} - x_{j} \right)^2 $$

We define

$$ \text{local var} (\textit{\textbf{x}}) = {\textit{\textbf{x}}}^T \textit{\textbf{L}} \textit{\textbf{x}}  \hspace{1cm}  \text{global var} (\textit{\textbf{x}}) = {\textit{\textbf{x}}}^T \left( \textit{\textbf{I}} - \frac {1} {n} {\textbf{1}}_{n \times n} \right) \textit{\textbf{x}}$$

Through some calculation and approximation, we could get the expression of global variance and local variance of potential and divergence.

$$\text{global var} (\textit{\textbf{p}}) \approx {\textit{\textbf{f}}}^T \textit{\textbf{U}}_{|\textit{E}| \times |\textit{E}|} \begin{bmatrix}
    {\bm{\Lambda}_{|\textit{V}| \times |\textit{V}|}}^{-1} & \textbf{0}_{|\textit{V}| \times (|\textit{E}|-|\textit{V}|)} \\
    \textbf{0}_{(|\textit{E}|-|\textit{V}|) \times |\textit{V}|} & \textbf{0}_{(|\textit{E}|-|\textit{V}|) \times (|\textit{E}|-|\textit{V}|)}
\end{bmatrix} {\textit{\textbf{U}}_{|\textit{E}| \times |\textit{E}|}}^T \textit{\textbf{f}} = \sum_{i=1}^{|\textit{V}|} \frac {a_i^2} {\lambda_i} $$

$$\text{global var} (\textit{\textbf{d}}) \approx {\textit{\textbf{f}}}^T \textit{\textbf{U}}_{|\textit{E}| \times |\textit{E}|} \begin{bmatrix}
    {\bm{\Lambda}_{|\textit{V}| \times |\textit{V}|}} & \textbf{0}_{|\textit{V}| \times (|\textit{E}|-|\textit{V}|)} \\
    \textbf{0}_{(|\textit{E}|-|\textit{V}|) \times |\textit{V}|} & \textbf{0}_{(|\textit{E}|-|\textit{V}|) \times (|\textit{E}|-|\textit{V}|)}
\end{bmatrix} {\textit{\textbf{U}}_{|\textit{E}| \times |\textit{E}|}}^T \textit{\textbf{f}} = \sum_{i=1}^{|\textit{V}|}  {\lambda_i}{a_i^2} $$

$$\text{local var} (\textit{\textbf{p}}) = {\textit{\textbf{f}}}^T \textit{\textbf{U}}_{|\textit{E}| \times |\textit{E}|} \begin{bmatrix}
    \textit{\textbf{I}}_{|\textit{V}| \times |\textit{V}|} & \textbf{0}_{|\textit{V}| \times (|\textit{E}|-|\textit{V}|)} \\
    \textbf{0}_{(|\textit{E}|-|\textit{V}|) \times |\textit{V}|} & \textbf{0}_{(|\textit{E}|-|\textit{V}|) \times (|\textit{E}|-|\textit{V}|)}
\end{bmatrix} {\textit{\textbf{U}}_{|\textit{E}| \times |\textit{E}|}}^T \textit{\textbf{f}} = \sum_{i=1}^{|\textit{V}|} {a_i^2} $$

$$\text{local var} (\textit{\textbf{d}}) = {\textit{\textbf{f}}}^T \textit{\textbf{U}}_{|\textit{E}| \times |\textit{E}|} \begin{bmatrix}
    {\bm{\Lambda}_{|\textit{V}| \times |\textit{V}|}}^2 & \textbf{0}_{|\textit{V}| \times (|\textit{E}|-|\textit{V}|)} \\
    \textbf{0}_{(|\textit{E}|-|\textit{V}|) \times |\textit{V}|} & \textbf{0}_{(|\textit{E}|-|\textit{V}|) \times (|\textit{E}|-|\textit{V}|)}
\end{bmatrix} {\textit{\textbf{U}}_{|\textit{E}| \times |\textit{E}|}}^T \textit{\textbf{f}}  = \sum_{i=1}^{|\textit{V}|} {\lambda_i}^2 {a_i^2} $$

$\textit{\textbf{U}}$, $\bm{\Lambda}$ and $\textit{\textbf{V}}$ are results from SVD of incidence matrix. $ \textit{\textbf{a}} = {\textit{\textbf{U}}_{|\textit{E}| \times |\textit{E}|}}^T \textit{\textbf{f}}$. These four values can all be expressed in a weighted sum of squares.

\begin{figure}[htbp]
    \centering
    \includegraphics[width=0.5\textwidth]{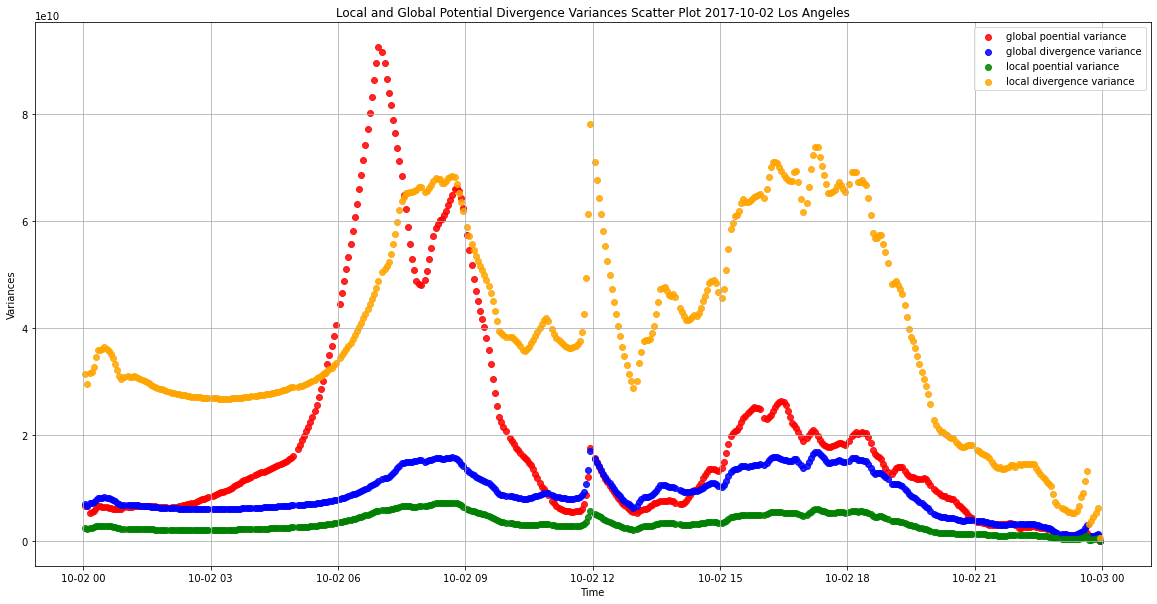}  
    \caption{The local and global variances for potential and divergence in a Day in LA}
    \label{fig:LA pot div var}
\end{figure}

We observe the properties of these four quantities are mostly determined by the singular values of the incidence matrix (or the eigenvalues of the graph Laplacian). From Appendix D, we can see that some eigenvalues of graph Laplacian are close to 0, while the majority of it is greater than 1. One reasonable explanation is that the squares of eigenvalues in the local variance of divergence make it a lot larger than the local variance of potential. And inverses of some eigenvalues very close to 0 make the global variance of potential larger than the global variance of divergence. The experiment results are shown in Figure \ref{fig:LA pot div var}.

\begin{figure}[htbp]
    \centering
    \includegraphics[width=0.5\textwidth]{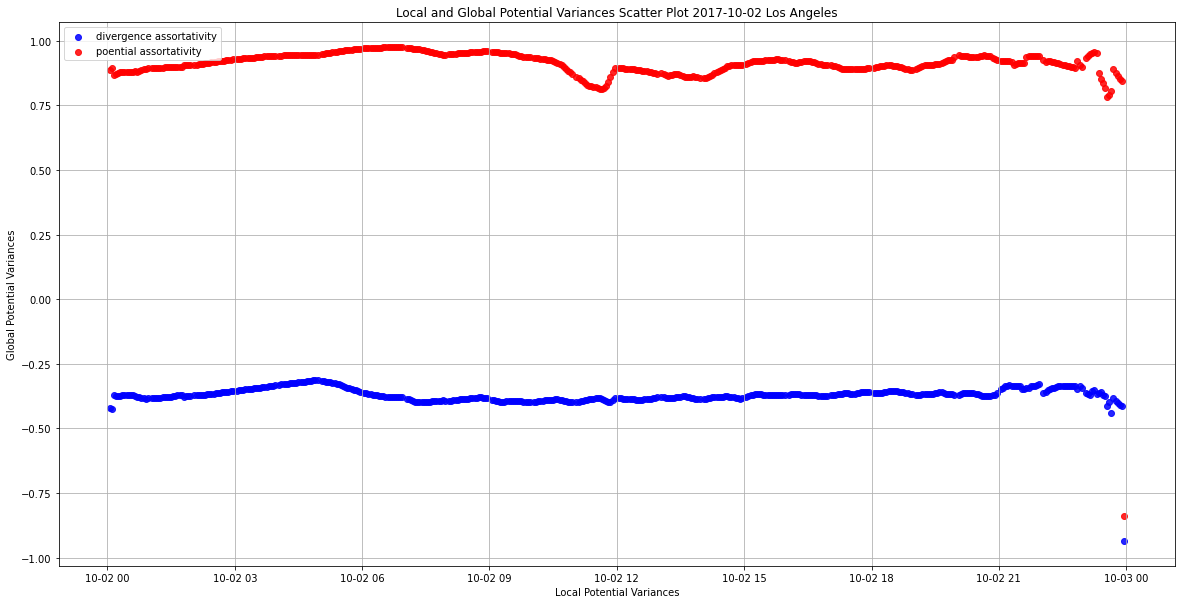}  
    \caption{The assortativities for potential and divergence in a Day in LA}
    \label{fig:LA assort}
\end{figure}

This tells us the potential is better than divergence in embedding nodes in the flow network. It is able to better differentiate different nodes globally and also embed neighboring nodes to similar values in the result. The assortativities of potential and divergence in Figure \ref{fig:LA assort} also show that potential values of neighboring nodes tend to be similar while divergence value of neighboring nodes tend to differ a lot.

Finally, we propose a new clustering scheme for flow graphs that combines three types of node‐level features:  
(1) structure: connectivity of the network (via symmetrized adjacency);  
(2) mean flow: average of flow volumes in both directions on each edge;  
(3) skew (net flow asymmetry): difference between directional flows.  

Formally, letting \(x_{ij}, x_{ji}\) be the flow volumes on edge \(i\!\to\!j\) and its reverse, define
\[
m_{ij} = \frac{x_{ij} + x_{ji}}{2},\qquad
s_{ij} = x_{ij} - x_{ji}.
\]
We then compute for each node \(i\) a feature vector
\[
Z_i = \Big[\,U_{\mathrm{struct}}(i,:),\; z_i^{(\mathrm{mean})},\; z_i^{(\mathrm{skew})} \Big]
\]
where \(U_{\mathrm{struct}}(i,:)\) are the first \(k\) eigenvectors of a Laplacian built from a symmetrized adjacency using \(m_{ij}\), and
\[
z_i^{(\mathrm{mean})} = \frac{1}{\deg_i}\sum_{j} w_{ij}\, m_{ij}, \quad
z_i^{(\mathrm{skew})} = \frac{1}{\deg_i}\sum_{j} w_{ij}\, s_{ij}.
\]
Clustering is then done via K-means on \(\{Z_i\}_{i\in V}\).  

We show in Appendix E that including the mean and skew components with the structural spectral embedding significantly improves cluster coherence over just using the structural embedding alone.

Note that standard Laplacian eigenmaps only use \(U_{\mathrm{struct}}(i,:)\) (i.e.\ from a Laplacian built purely from undirected / symmetrized adjacency), whereas our method additionally incorporates 1-dimensional scalar features \(z^{(\mathrm{mean})}\) and \(z^{(\mathrm{skew})}\) which encode flow magnitude and direction, respectively.

\section*{Discussion}

Analysis of properties of potential and divergence reveals possible applications in clustering and embedding in directed flow graphs. More studies could be done on how to define similarity and better cluster in flow graphs. If there is simultaneous bidirectional flow on the edges, considering using both the mean and difference of two flows could be better to incorporate all information on those edges. \cite{Schaub2021HONtutorial}

Due to the limited amount of publicly available road-level real-time traffic flow dataset. We could not test the results on more cities. Note that the road network in other UTD19 dataset cities are either incomplete or small or disconnected, which makes the results to be not useful on any of them.

However, a drawback of this potential metric is that computing this involves doing a pseudo-inverse of the graph Laplacian, which (with a dense direct method) incurs $O(n^3)$ time complexity. In theory, one can express the potential \(p\) by using the pseudo-inverse of the Laplacian:
\[
p = L^{\dagger} d,
\]
which corresponds to “inverting” \(L\) (ignoring its nullspace) and thus suggests an \(O(n^3)\) complexity if done naively via dense matrix factorization. However, in practice \(L\) is often sparse (graphs with \(m \ll n^2\) edges), and one can instead solve the system
\[
L\,p = d
\]
using sparse direct solvers (e.g. sparse Cholesky) or iterative methods (e.g. Conjugate Gradient, PCG) with appropriate preconditioning, which scale approximately as \(O(m \log^c n)\) or in favorable cases nearly linear in \(m\).   In practice, the Laplacian matrix is “sparse” (number of edges $m \ll n^2$), and one may employ “nearly-linear time solvers” for symmetric diagonally dominant (SDD) systems, such as the Spielman–Teng solver \citep{SpielmanTeng2004, SpielmanTeng2006}, or algebraic multigrid solvers such as LAMG \citep{LivneBrandt2011}.  These methods can solve $Lx = b$ in time approximately $O(m \log^{c} n)$ or even $O(m)$ in favorable cases.

\section*{Conclusion}
We have provided empirical evidence / analysis shows that HodgeRank does not give interesting results when applied in an Origin-Destination network. This is because the very dense structure of the graph causes the graph laplacian to be close to a diagonal matrix. This diagonal matrix makes the relationship between potential and divergence to be very linear. The linear coefficients are close to degrees of nodes. However, in a mesh-like road network, potential shares lower similarity to divergence. It provides a smoother but more distinguishable metric for the nodes in the graph. We could observe obvious flow direction reversal in the potential graph in the morning and evening rush hours. Finally, I propose that in a flow graph, we could add the potential information extracted from edge flows in addition to the common graph laplacian, which only encodes the topology of the graph by connectivity, to better do clustering on the graph.

\newpage

\nocite{*}
\bibliographystyle{plainnat} 
\bibliography{ref}           

\begin{thebibliography}{28}
\providecommand{\natexlab}[1]{#1}
\providecommand{\url}[1]{\texttt{#1}}
\expandafter\ifx\csname urlstyle\endcsname\relax
  \providecommand{\doi}[1]{doi: #1}\else
  \providecommand{\doi}{doi: \begingroup \urlstyle{rm}\Url}\fi

\bibitem[Aoki et~al.(2022)Aoki, Fujishima, and Fujiwara]{aoki2022urban}
Takaaki Aoki, Shota Fujishima, and Naoya Fujiwara.
\newblock Urban spatial structures from human flow by hodge--kodaira decomposition.
\newblock \emph{Scientific reports}, 12\penalty0 (1):\penalty0 11258, 2022.

\bibitem[Arnold et~al.(2006)Arnold, Falk, and Winther]{ArnoldFalkWinther2006FEEC}
Douglas~N. Arnold, Richard~S. Falk, and Ragnar Winther.
\newblock Finite element exterior calculus: from hodge theory to numerical stability.
\newblock \emph{Acta Numerica}, 15:\penalty0 1--155, 2006.
\newblock \doi{10.1017/S0962492906210018}.

\bibitem[Belkin and Niyogi(2003)]{BelkinNiyogi2003}
Mikhail Belkin and Partha Niyogi.
\newblock Laplacian eigenmaps for dimensionality reduction and data representation.
\newblock \emph{Neural Computation}, 15\penalty0 (6):\penalty0 1373--1396, 2003.
\newblock \doi{10.1162/089976603321780317}.

\bibitem[Chung(1997)]{Chung1997}
Fan R.~K. Chung.
\newblock \emph{Spectral Graph Theory}, volume~92 of \emph{CBMS Regional Conference Series in Mathematics}.
\newblock American Mathematical Society, 1997.

\bibitem[Desbrun et~al.(2005)Desbrun, Hirani, Leok, and Marsden]{DesbrunHiraniLeokMarsden2005DEC}
Mathieu Desbrun, Anil~N. Hirani, Melvin Leok, and Jerrold~E. Marsden.
\newblock Discrete exterior calculus, 2005.
\newblock Tutorial/technical report.

\bibitem[{ETH Zurich}(2019{\natexlab{a}})]{UTD19-manual}
{ETH Zurich}.
\newblock Utd19 manual and citation.
\newblock \url{https://utd19.ethz.ch/docs/manual-and-citation}, 2019{\natexlab{a}}.
\newblock Dataset manual and citation instructions.

\bibitem[{ETH Zurich}(2019{\natexlab{b}})]{UTD19-web}
{ETH Zurich}.
\newblock Utd19: Urban traffic data set 2019.
\newblock \url{https://utd19.ethz.ch/}, 2019{\natexlab{b}}.
\newblock Project page with dataset description, 40 cities, 15-min resolution, license and citation.

\bibitem[Fiedler(1973)]{Fiedler1973}
Miroslav Fiedler.
\newblock Algebraic connectivity of graphs.
\newblock \emph{Czechoslovak Mathematical Journal}, 23\penalty0 (98):\penalty0 298--305, 1973.

\bibitem[Hagen and Kahng(1992)]{HagenKahng1992}
Lars Hagen and Andrew~B. Kahng.
\newblock New spectral methods for ratio cut partitioning and clustering.
\newblock \emph{IEEE Transactions on Computer-Aided Design of Integrated Circuits and Systems}, 11\penalty0 (9):\penalty0 1074--1085, 1992.
\newblock \doi{10.1109/43.159993}.

\bibitem[Horak and Jost(2013)]{HorakJost2013}
Danijela Horak and J{\"u}rgen Jost.
\newblock Spectra of combinatorial laplace operators on simplicial complexes.
\newblock \emph{Advances in Mathematics}, 244:\penalty0 303--336, 2013.
\newblock \doi{10.1016/j.aim.2013.05.011}.

\bibitem[Hu et~al.(2022)Hu, Shao, and Sun]{electronics11152432}
Zhiqiu Hu, Fengjing Shao, and Rencheng Sun.
\newblock A new perspective on traffic flow prediction: A graph spatial-temporal network with complex network information.
\newblock \emph{Electronics}, 11\penalty0 (15), 2022.
\newblock ISSN 2079-9292.
\newblock URL \url{https://www.mdpi.com/2079-9292/11/15/2432}.

\bibitem[Ji()]{zj4442022}
Zhao Ji.
\newblock Hodgerank: Generating movie ranking from imdb movie ratings, part 1.
\newblock URL \url{https://medium.com/@zj444/hodgerank-generating-movie-ranking-from-imdb-movie-ratings-part-1-2a88ec148f10}.
\newblock Accessed: April 16, 2024.

\bibitem[Jiang et~al.(2011)Jiang, Lim, Yao, and Ye]{jiang2011statistical}
Xiaoye Jiang, Lek-Heng Lim, Yuan Yao, and Yinyu Ye.
\newblock Statistical ranking and combinatorial hodge theory.
\newblock \emph{Mathematical Programming}, 127\penalty0 (1):\penalty0 203--244, 2011.

\bibitem[Lim(2020)]{lim2020hodge}
Lek-Heng Lim.
\newblock Hodge laplacians on graphs.
\newblock \emph{Siam Review}, 62\penalty0 (3):\penalty0 685--715, 2020.

\bibitem[Livne and Brandt(2011)]{LivneBrandt2011}
Oren~E. Livne and Achi Brandt.
\newblock Lean algebraic multigrid (lamg): Fast graph laplacian linear solver.
\newblock \emph{arXiv preprint arXiv:1108.0123}, 2011.

\bibitem[Loder et~al.(2019)Loder, Amb{\"u}hl, Menendez, and Axhausen]{LoderAmbuehlMenendezAxhausen2019}
Allister Loder, Lukas Amb{\"u}hl, Monica Menendez, and Kay W. et~al. Axhausen.
\newblock Understanding traffic capacity of urban networks.
\newblock \emph{Scientific Reports}, 9\penalty0 (1):\penalty0 16283, 2019.
\newblock \doi{10.1038/s41598-019-51539-5}.

\bibitem[Ng et~al.(2002)Ng, Jordan, and Weiss]{NgJordanWeiss2002}
Andrew~Y. Ng, Michael~I. Jordan, and Yair Weiss.
\newblock On spectral clustering: Analysis and an algorithm.
\newblock In \emph{Advances in Neural Information Processing Systems (NIPS)}, pages 849--856, 2002.

\bibitem[{Puget Sound Regional Council}(2022)]{PSRC2021Hub}
{Puget Sound Regional Council}.
\newblock Household travel survey 2021 — psrc arcgis hub, 2022.
\newblock URL \url{https://psrc-psrc.hub.arcgis.com/datasets/psrc::household-travel-survey-2021/about}.
\newblock Accessed: 2025-09-21.

\bibitem[Schaub et~al.(2021)Schaub, Zhu, S{\'e}by, Roddenberry, and Segarra]{Schaub2021HONtutorial}
Michael~T. Schaub, Yu~Zhu, Jean-Baptiste S{\'e}by, T.~Mitchell Roddenberry, and Santiago Segarra.
\newblock Signal processing on higher-order networks: Livin’ on the edge … and beyond.
\newblock \emph{arXiv preprint arXiv:2101.05510}, 2021.
\newblock URL \url{https://arxiv.org/abs/2101.05510}.

\bibitem[Shi and Malik(2000)]{ShiMalik2000}
Jianbo Shi and Jitendra Malik.
\newblock Normalized cuts and image segmentation.
\newblock \emph{IEEE Transactions on Pattern Analysis and Machine Intelligence}, 22\penalty0 (8):\penalty0 888--905, 2000.
\newblock \doi{10.1109/34.868688}.

\bibitem[Sol{\'e}-Ribalta et~al.(2016)Sol{\'e}-Ribalta, G{\'o}mez, and Arenas]{sole2016model}
Albert Sol{\'e}-Ribalta, Sergio G{\'o}mez, and Alex Arenas.
\newblock A model to identify urban traffic congestion hotspots in complex networks.
\newblock \emph{Royal Society open science}, 3\penalty0 (10):\penalty0 160098, 2016.

\bibitem[Spielman and Teng(2004)]{SpielmanTeng2004}
Daniel~A. Spielman and Shang-Hua Teng.
\newblock Nearly-linear time algorithms for random electrical networks.
\newblock \emph{Proceedings of the 36th Annual ACM Symposium on Theory of Computing (STOC)}, pages 81--90, 2004.

\bibitem[Spielman and Teng(2006)]{SpielmanTeng2006}
Daniel~A. Spielman and Shang-Hua Teng.
\newblock Nearly-linear time algorithms for preconditioning and solving symmetric, diagonally dominant linear systems.
\newblock \emph{SIAM Journal on Matrix Analysis and Applications}, 35\penalty0 (3):\penalty0 835--885, 2006.

\bibitem[{U.S. Census Bureau}(2023)]{OnTheMap}
{U.S. Census Bureau}.
\newblock Onthemap: Lehd origin-destination employment statistics.
\newblock \url{https://onthemap.ces.census.gov/}, 2023.
\newblock Official portal for LODES-based OD analyses.

\bibitem[{U.S. Census Bureau, LEHD Program}(2023)]{LEHD-LODES-techdoc}
{U.S. Census Bureau, LEHD Program}.
\newblock Lehd origin-destination employment statistics (lodes) technical documentation.
\newblock \url{https://lehd.ces.census.gov/data/}, 2023.
\newblock Technical definitions, methods, and data products.

\bibitem[{U.S. Census Bureau, LEHD Program}(2024)]{LEHDLODESOverview}
{U.S. Census Bureau, LEHD Program}.
\newblock Lehd origin-destination employment statistics (lodes): Product overview, 2024.
\newblock URL \url{https://lehd.ces.census.gov/data/lodes/LODES7/LODES7_Tech_Documentation.pdf}.
\newblock General overview page and documentation links; Accessed: 2025-09-21.

\bibitem[von Luxburg(2007)]{vonLuxburg2007tutorial}
Ulrike von Luxburg.
\newblock A tutorial on spectral clustering.
\newblock \emph{Statistics and Computing}, 17\penalty0 (4):\penalty0 395--416, 2007.
\newblock \doi{10.1007/s11222-007-9033-z}.

\bibitem[Zhang et~al.(2022)Zhang, Huang, Guo, and He]{ZHANG2022128063}
Mengyao Zhang, Tao Huang, Zhaoxia Guo, and Zhenggang He.
\newblock Complex-network-based traffic network analysis and dynamics: A comprehensive review.
\newblock \emph{Physica A: Statistical Mechanics and its Applications}, 607:\penalty0 128063, 2022.
\newblock ISSN 0378-4371.
\newblock \doi{https://doi.org/10.1016/j.physa.2022.128063}.
\newblock URL \url{https://www.sciencedirect.com/science/article/pii/S0378437122006628}.

\end{thebibliography}

\newpage

\section*{Appendix A}

For further inquiries or to access additional resources, please use the following contact information and links:

- Email: yifeisun@umich.edu

- GitHub Repository: \href{https://github.com/YifeiSun01/HodgeRank}{https://github.com/YifeiSun01/HodgeRank}

\newpage

\section*{Appendix B}

\begin{figure}[htbp]
  \centering
  \begin{subfigure}[b]{0.4\textwidth}
    \includegraphics[width=\textwidth]{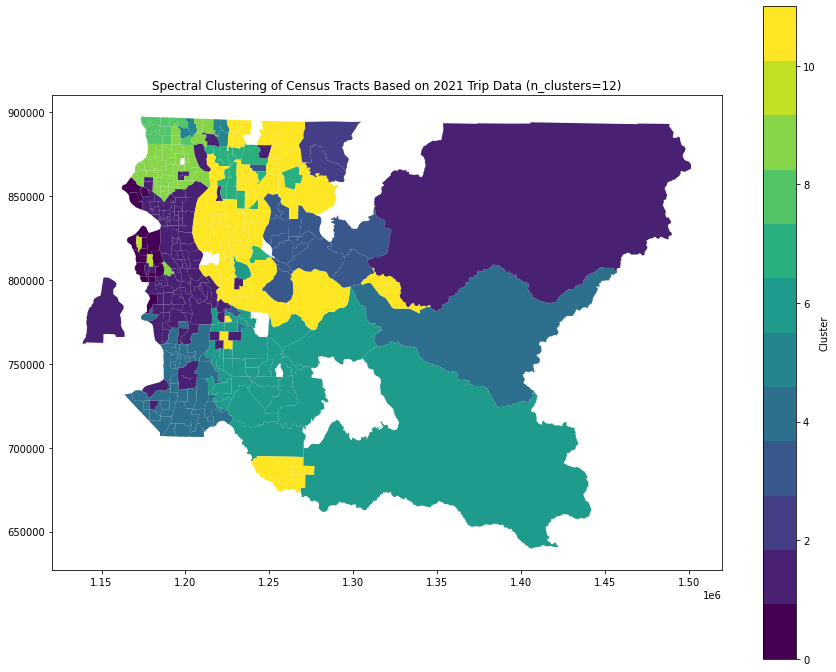}
    \caption{Spectral Clustering of Census Tracts in King County, Washington Based on PSRC Data}
    \label{fig:king county clustering 1}
  \end{subfigure}
  \hfill
  \begin{subfigure}[b]{0.4\textwidth}
    \includegraphics[width=\textwidth]{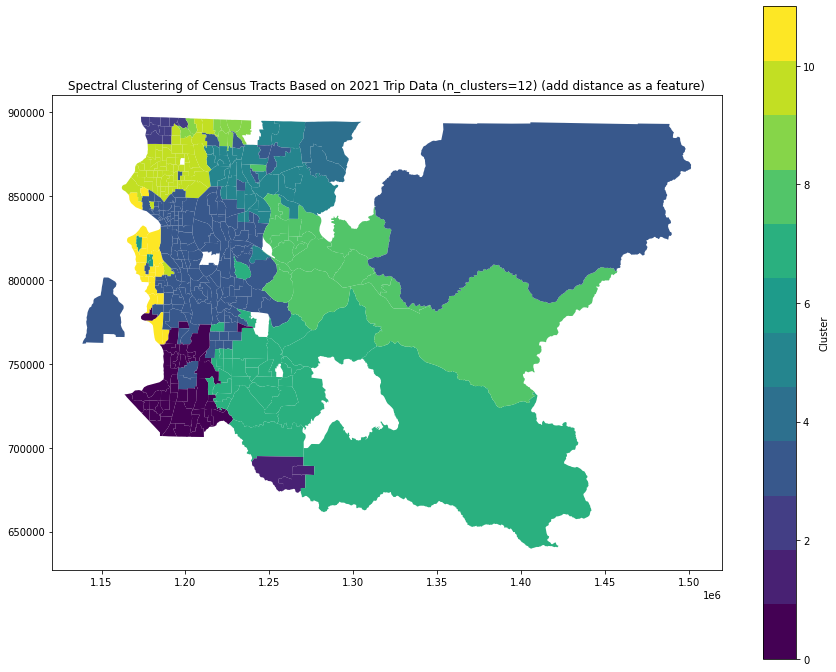}
    \caption{Spectral Clustering of Census Tracts in King County, Washington Based on PSRC Data and Distance Data}
    \label{fig:king county clustering 2}
  \end{subfigure}
  \begin{subfigure}[b]{0.4\textwidth}
    \includegraphics[width=\textwidth]{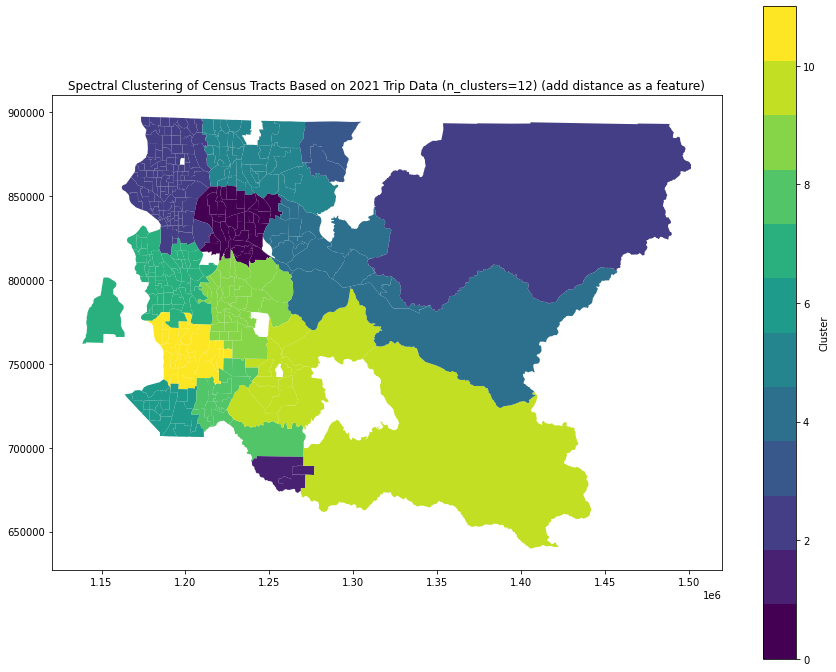}
    \caption{Spectral Clustering of Census Tracts in King County, Washington Based on PSRC Data and Distance Data (Larger Weight for Distance)}
    \label{fig:king county clustering 3}
  \end{subfigure}
  \caption{Spectral Clustering of Census Tracts in King County, Washington Based on PSRC Data}
  \label{fig:king county clustering}
\end{figure}

\begin{figure}[htbp]
  \centering
  \begin{subfigure}[b]{0.4\textwidth}
    \includegraphics[width=\textwidth]{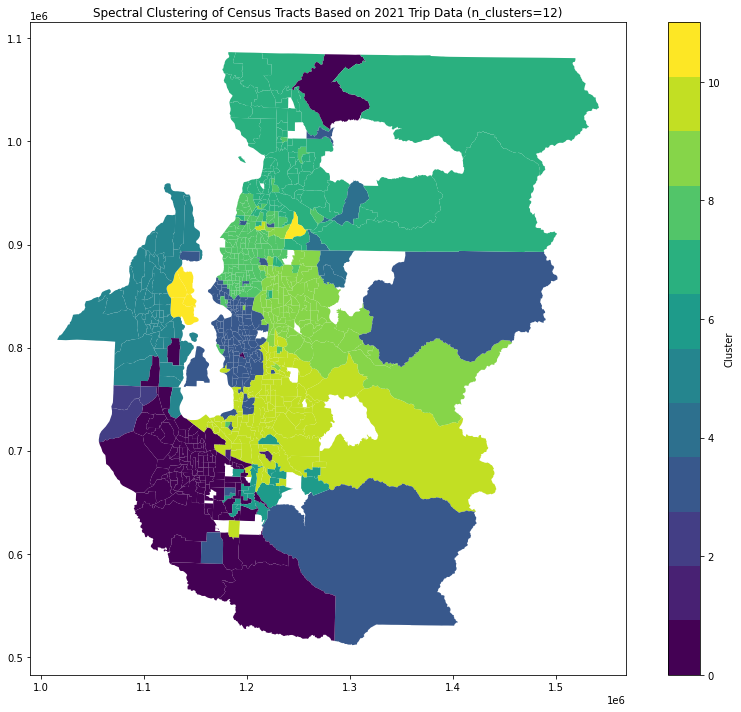}
    \caption{Spectral Clustering of Census Tracts in Puget Sound Based on PSRC Data}
    \label{fig:puget sound clustering 1}
  \end{subfigure}
  \hfill
  \begin{subfigure}[b]{0.4\textwidth}
    \includegraphics[width=\textwidth]{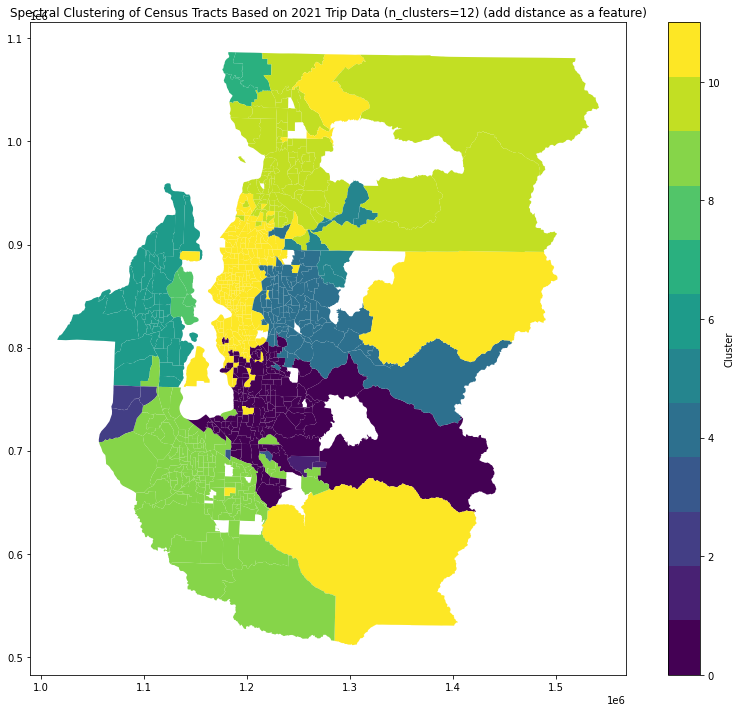}
    \caption{Spectral Clustering of Census Tracts in Puget Sound Based on PSRC Data and Distance Data}
    \label{fig:puget sound clustering 2}
  \end{subfigure}
  \begin{subfigure}[b]{0.4\textwidth}
    \includegraphics[width=\textwidth]{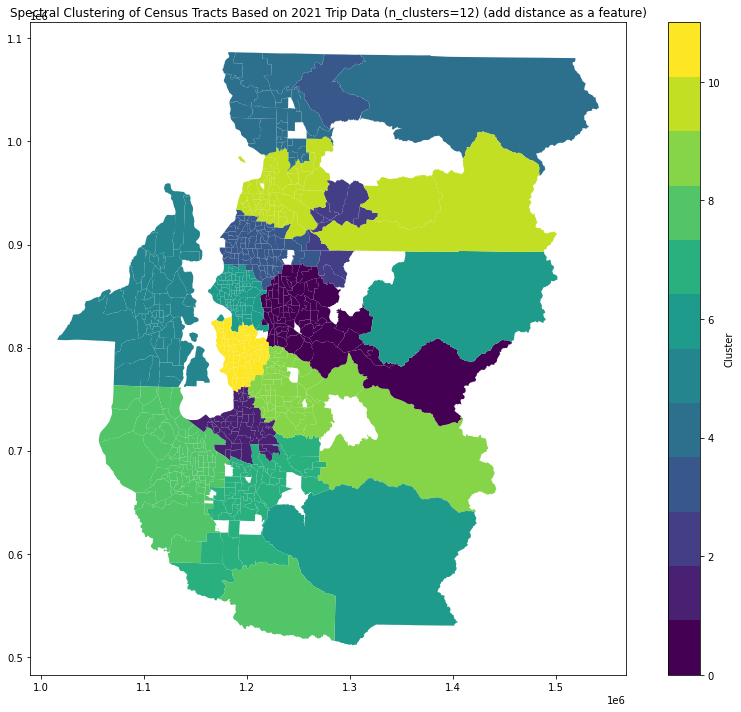}
    \caption{Spectral Clustering of Census Tracts in Puget Sound Based on PSRC Data and Distance Data (Larger Weight for Distance)}
    \label{fig:puget sound clustering 3}
  \end{subfigure}
  \caption{Spectral Clustering of Census Tracts in Puget Sound Based on PSRC Data}
  \label{fig:puget sound clustering}
\end{figure}

\newpage

\section*{Appendix C}

\begin{figure}[htbp]
  \centering
  \begin{subfigure}[b]{0.4\textwidth}
    \includegraphics[width=\textwidth]{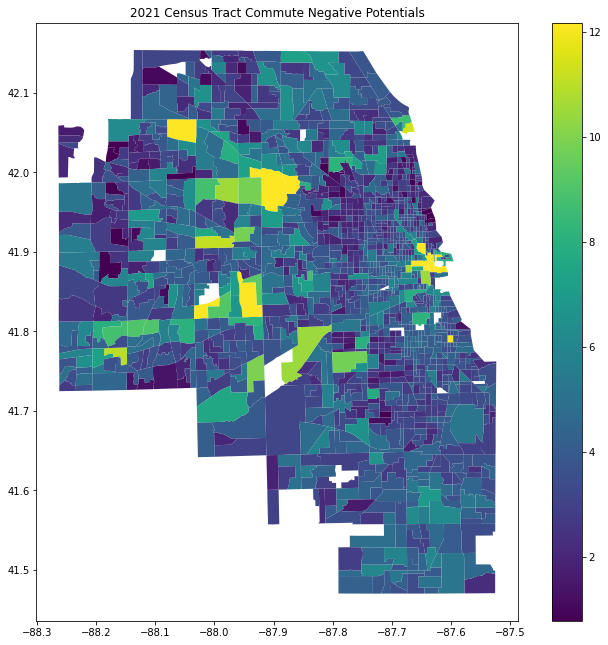}
    \caption{Negative Potentials of Census Tracts in Chicago Metro Area}
    \label{fig:chicago pot}
  \end{subfigure}
  \hfill
  \begin{subfigure}[b]{0.4\textwidth}
    \includegraphics[width=\textwidth]{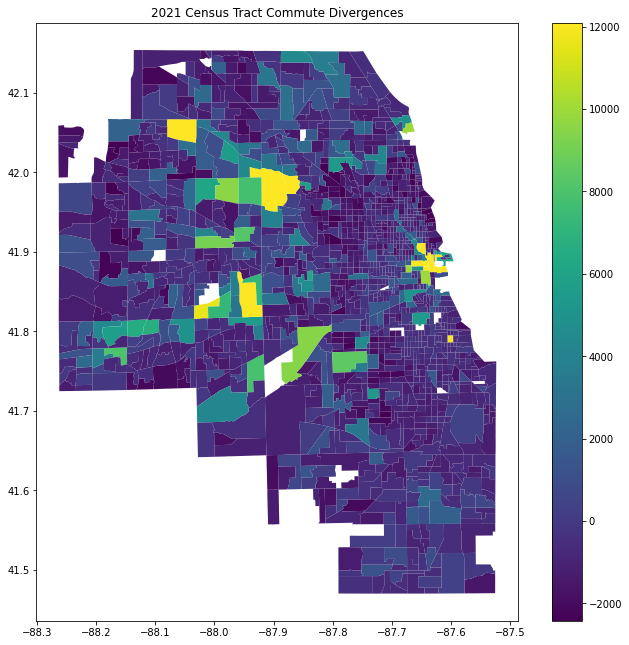}
    \caption{Divergences of Census Tracts in Chicago Metro Area}
    \label{fig:chicago div}
  \end{subfigure}
  \begin{subfigure}[b]{0.4\textwidth}
    \includegraphics[width=\textwidth]{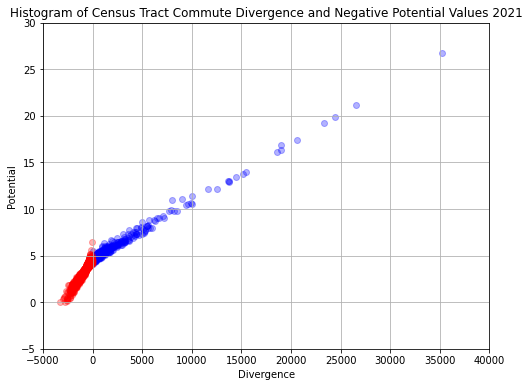}
    \caption{Negative Potentials and Divergences of Census Tracts in Chicago Metro Area}
    \label{fig:chicago pot div}
  \end{subfigure}
  \begin{subfigure}[b]{0.4\textwidth}
    \includegraphics[width=\textwidth]{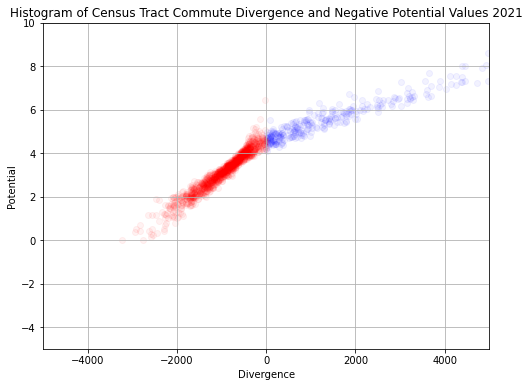}
    \caption{Negative Potentials and Divergences of Census Tracts in Chicago Metro Area (Enlarged)}
    \label{fig:chicago pot div large}
  \end{subfigure}
  \caption{Negative Potential and Divergence in Chicago Metro Area}
  \label{fig:Chicago}
\end{figure}

\begin{figure}[htbp]
  \centering
  \begin{subfigure}[b]{0.4\textwidth}
    \includegraphics[width=\textwidth]{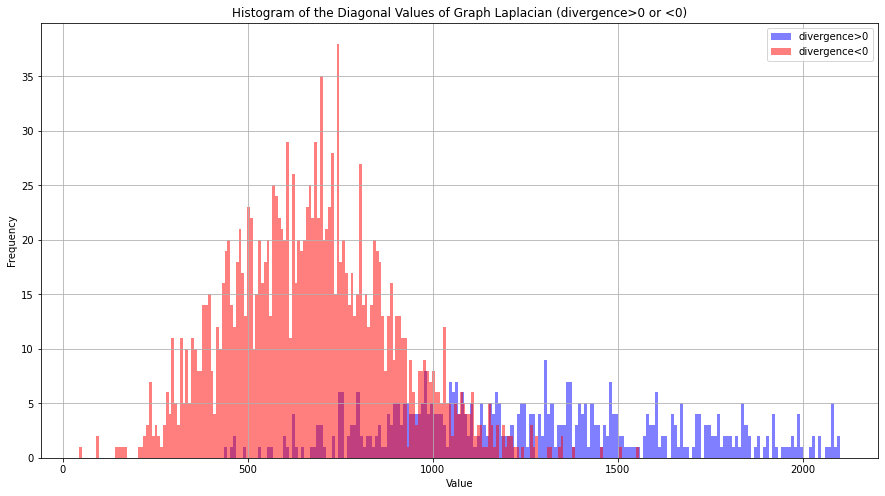}
    \caption{Graph Laplacian Diagonal Values}
    \label{fig:chicago graph laplacian diagonal}
  \end{subfigure}
  \begin{subfigure}[b]{0.4\textwidth}
    \includegraphics[width=\textwidth]{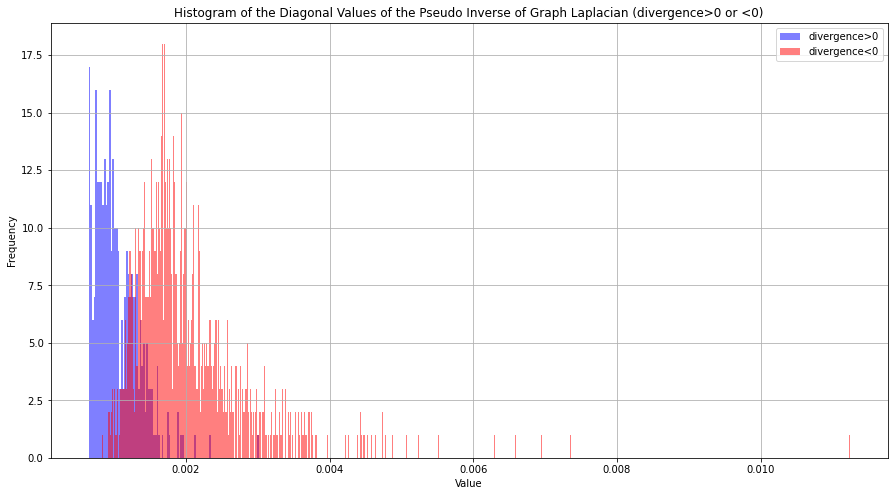}
    \caption{Graph Laplacian Pseudo Inverse Diagonal Values}
    \label{fig:chicago pseudo inverse diagonal}
  \end{subfigure}
  \begin{subfigure}[b]{0.4\textwidth}
    \includegraphics[width=\textwidth]{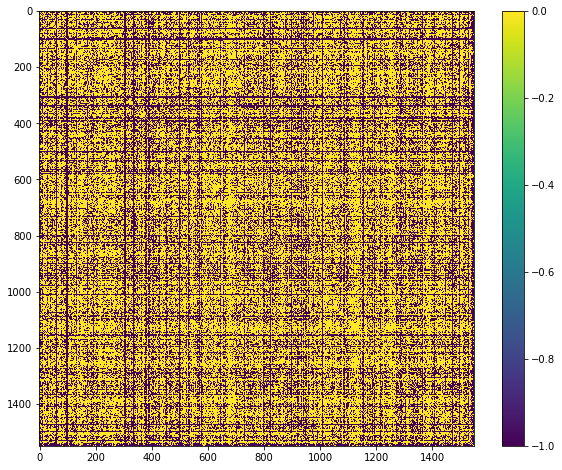}
    \caption{Graph Laplacian Non-Diagonal Values}
    \label{fig:chicago graph laplacian non diagonal}
  \end{subfigure}
  \begin{subfigure}[b]{0.4\textwidth}
    \includegraphics[width=\textwidth]{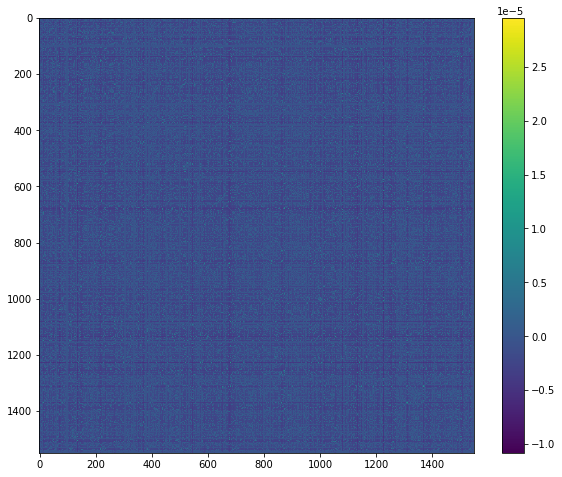}
    \caption{Graph Laplacian Pseudo Inverse Non-Diagonal Values}
    \label{fig:chicago pseudo inverse non diagonal}
  \end{subfigure}
  \caption{Negative Potential and Divergence in Chicago Metro Area}
  \label{fig:Chicago}
\end{figure}

\newpage

\section*{Appendix D}

\begin{figure}[htbp]
  \centering
    \includegraphics[width=0.4\textwidth]{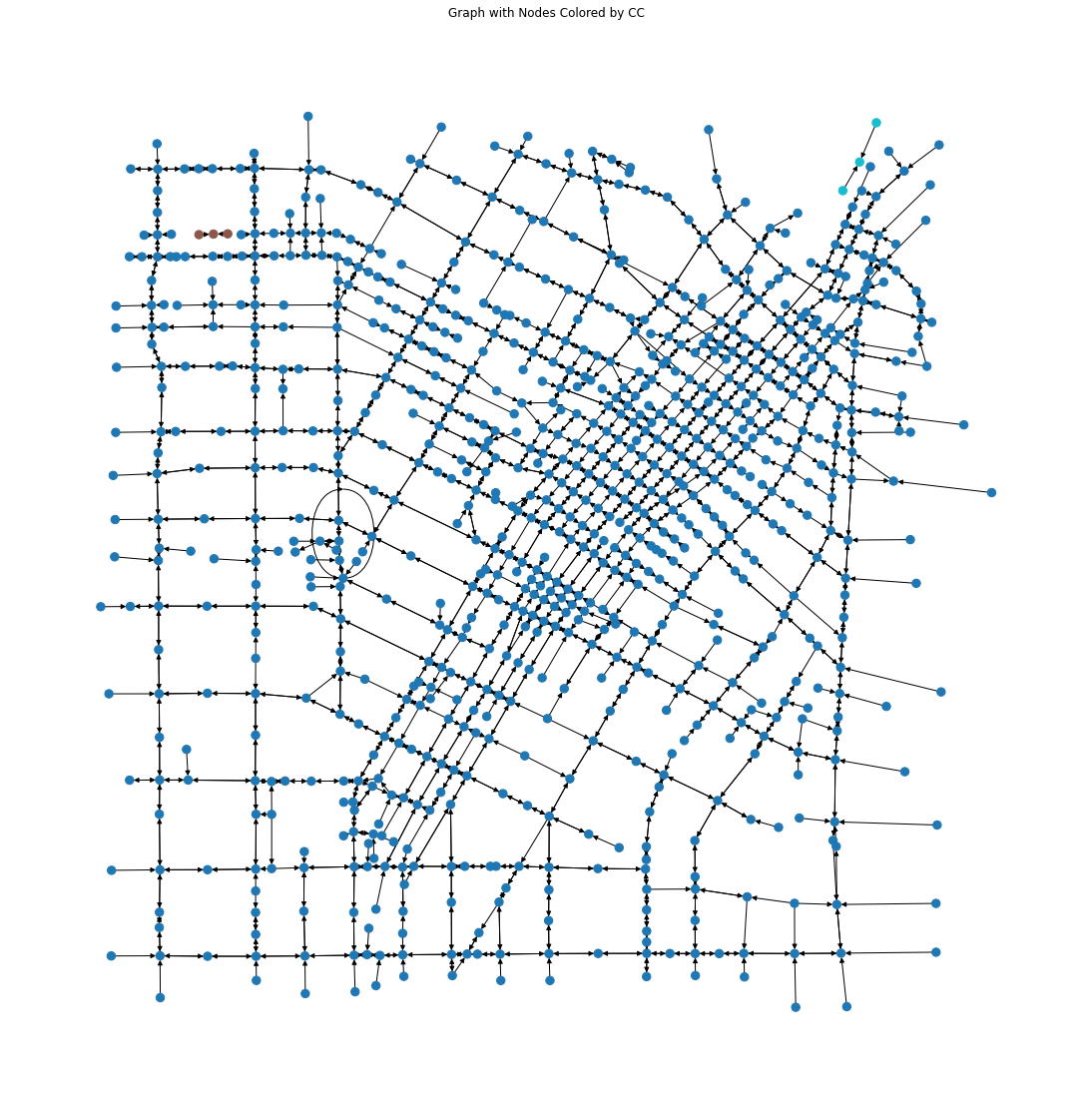}  
    \label{fig:LA network}
  \caption{Road Network of Downtown LA}
  \label{fig:LA Road Network}
\end{figure}

\begin{figure}[htbp]
  \centering
  \begin{subfigure}[b]{0.4\textwidth}
    \includegraphics[width=\textwidth]{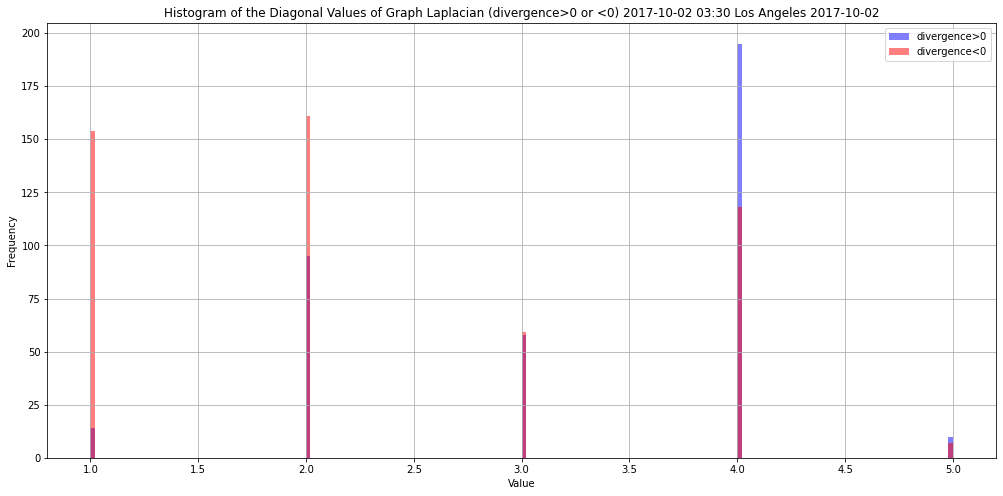}
    \caption{Degree Distributions of Road Network of Downtown LA}
    \label{fig:LA graph laplacian diagonal}
  \end{subfigure}
  \begin{subfigure}[b]{0.4\textwidth}
    \includegraphics[width=\textwidth]{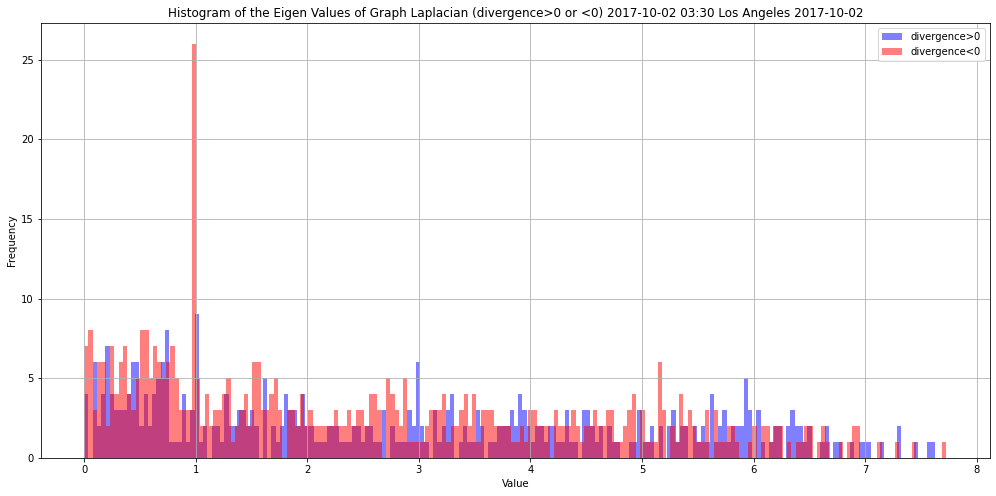}
    \caption{Eigenvalue Distribution of Graph Laplacian of Road Network of Downtown LA}
    \label{fig:LA graph laplacian eigenvalues}
  \end{subfigure}
  \begin{subfigure}[b]{0.4\textwidth}
    \includegraphics[width=\textwidth]{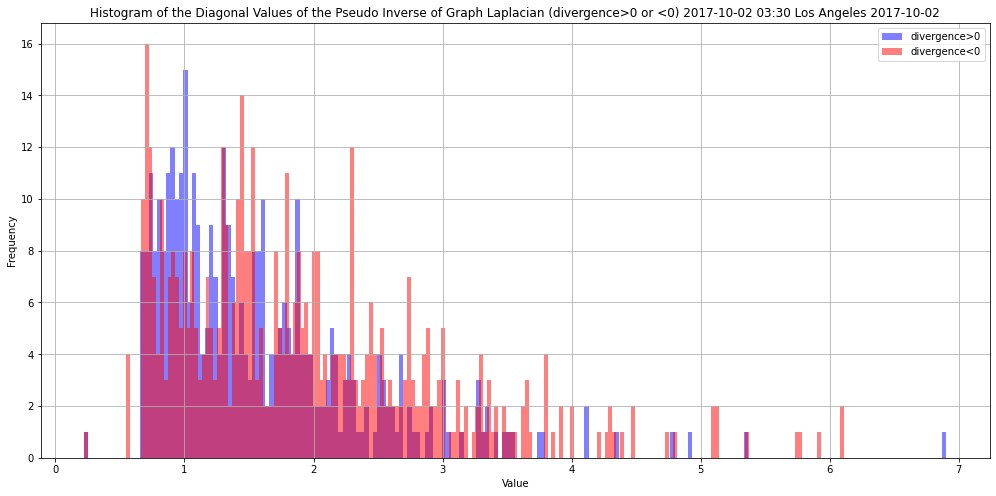}
    \caption{Diagonal Values of Pseudo Inverse of Graph Laplacian of Road Network of Downtown LA}
    \label{fig:LA pseudo inverse diagonal}
  \end{subfigure}
  \begin{subfigure}[b]{0.4\textwidth}
    \includegraphics[width=\textwidth]{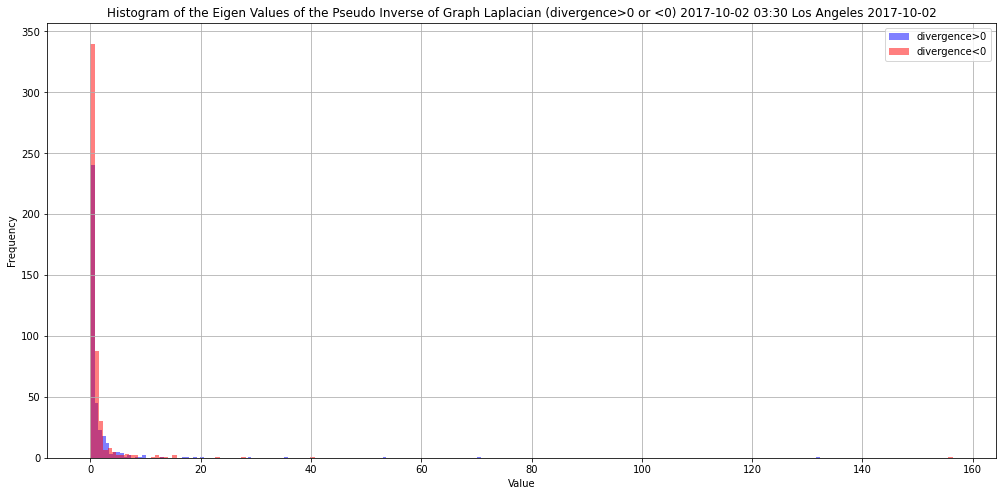}
    \caption{Eigenvalue Distribution of Pseudo Inverse of Graph Laplacian of Road Network of Downtown LA}
    \label{fig:LA pseudo inverse eigenvalues}
  \end{subfigure}
      \caption{Properties of Graph Laplacian of Road Network of Downtown LA}
  \label{fig:LA hodge 8:30}
\end{figure}

\begin{figure}[htbp]
  \centering
  \begin{subfigure}[b]{0.4\textwidth}
    \includegraphics[width=\textwidth]{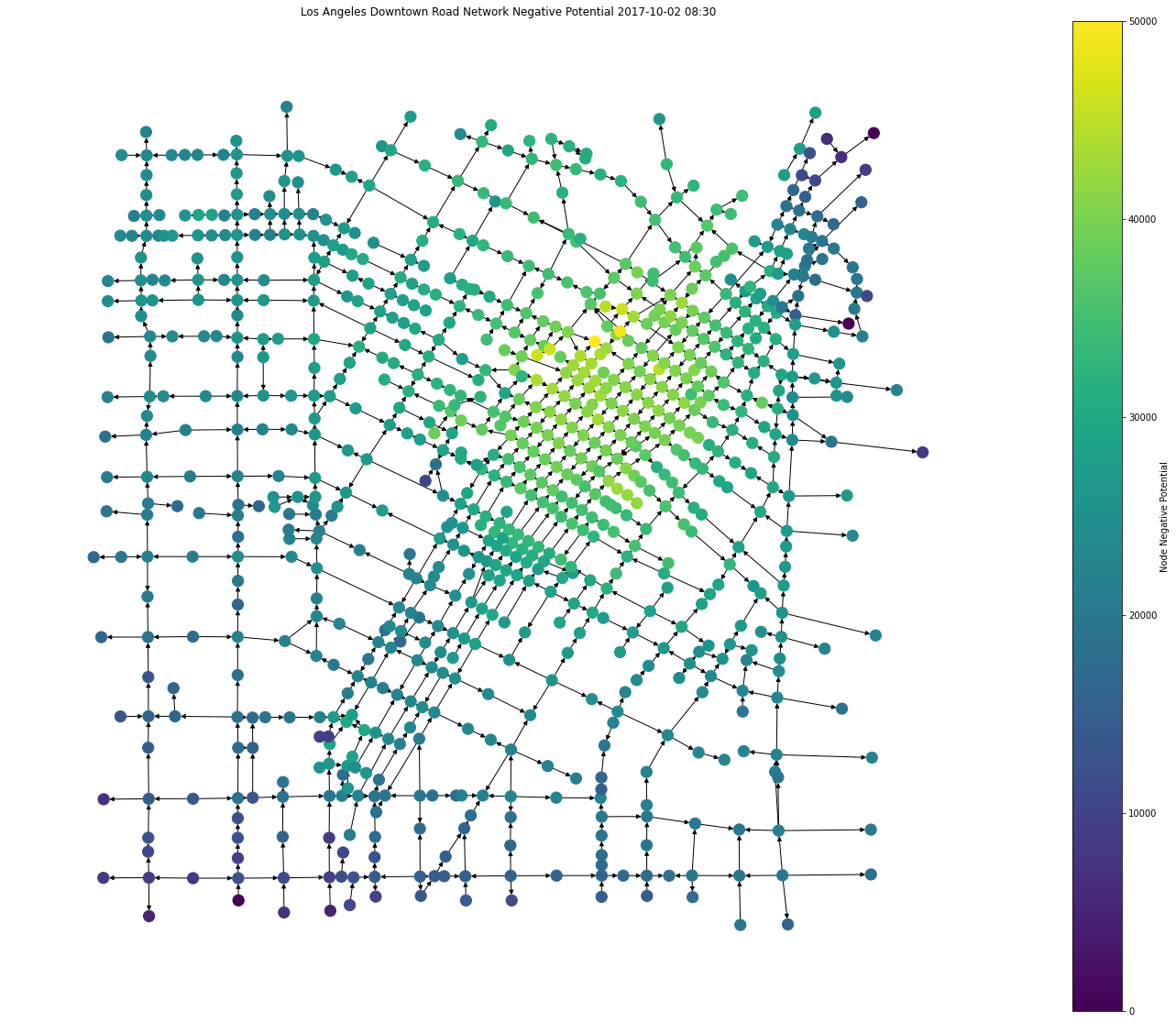}
    \caption{Potential in Road Network of Downtown LA at 8:30}
    \label{fig:LA 8 potential}
  \end{subfigure}
  \begin{subfigure}[b]{0.4\textwidth}
    \includegraphics[width=\textwidth]{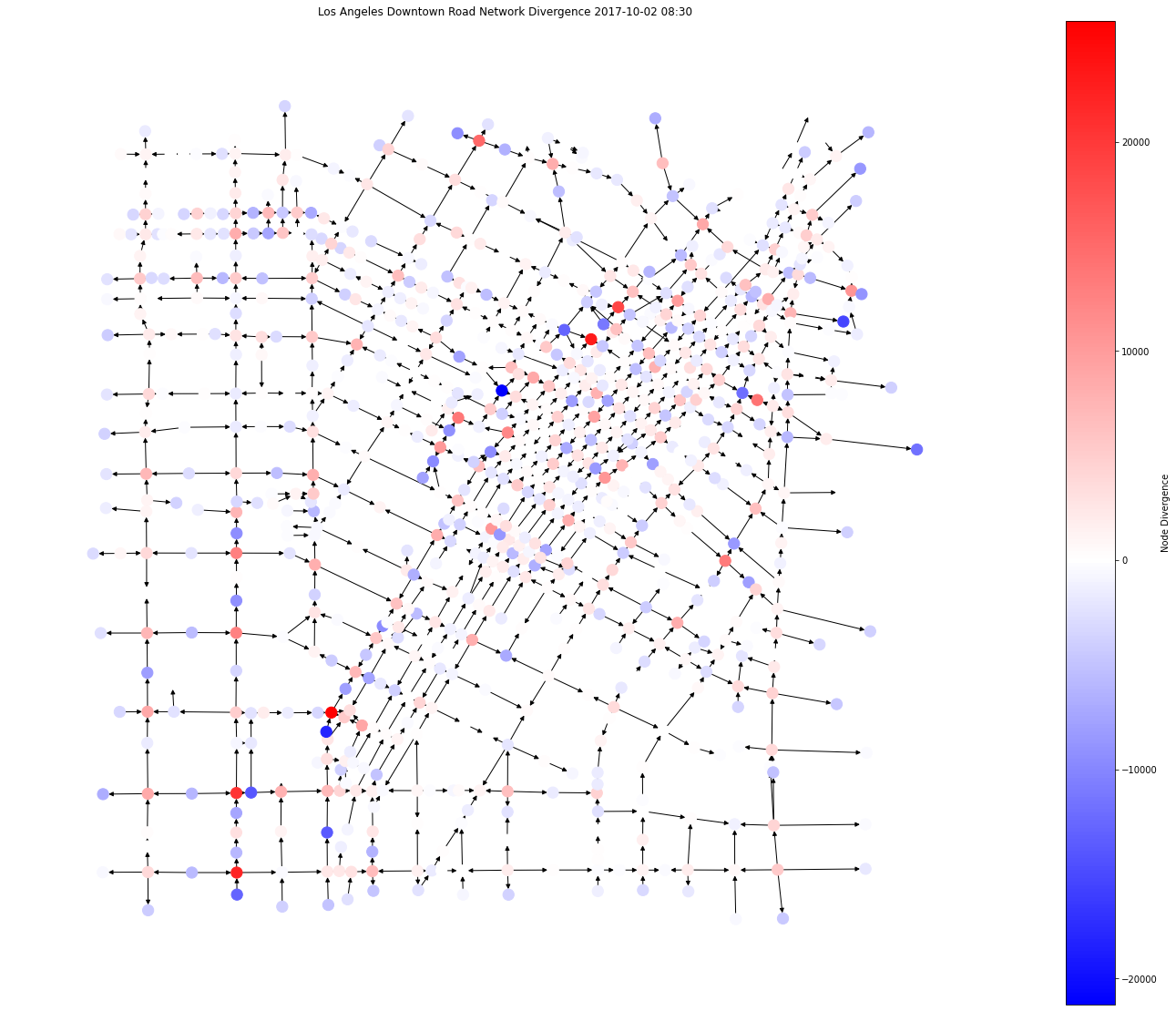}
    \caption{Divergence in Road Network of Downtown LA at 8:30}
    \label{fig:LA 8 divergence}
  \end{subfigure}
  \begin{subfigure}[b]{0.4\textwidth}
    \includegraphics[width=\textwidth]{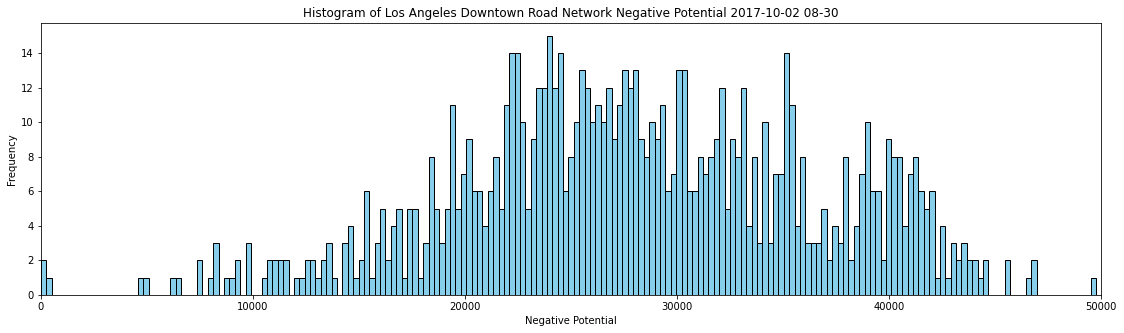}
    \caption{Potential Distribution in Road Network of Downtown LA at 8:30}
    \label{fig:LA 8 pot dist}
  \end{subfigure}
  \begin{subfigure}[b]{0.4\textwidth}
    \includegraphics[width=\textwidth]{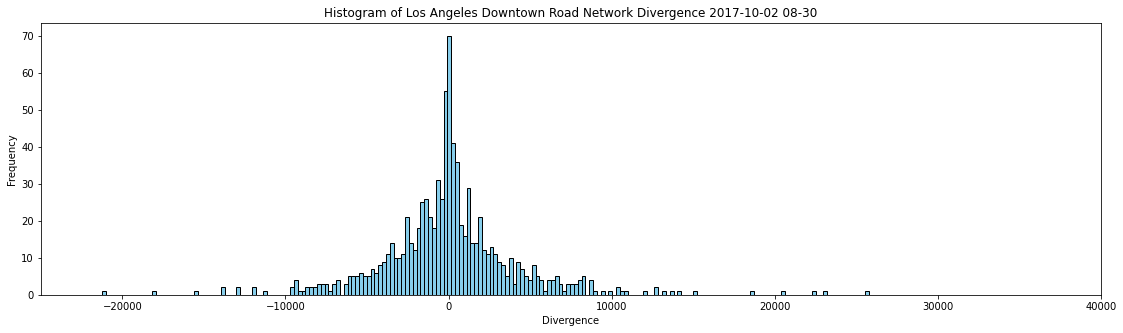}
    \caption{Divergence Distribution in Road Network of Downtown LA at 8:30}
    \label{fig:LA 8 div dist}
  \end{subfigure}
  \begin{subfigure}[b]{0.4\textwidth}
    \includegraphics[width=\textwidth]{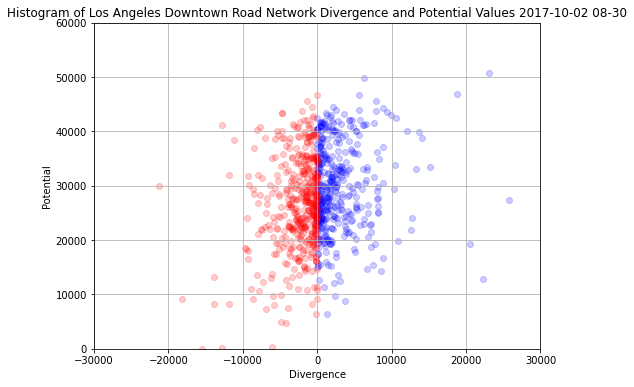}
    \caption{Potential and Divergence in Road Network of Downtown LA at 8:30}
    \label{fig:LA 8 pot div}
  \end{subfigure}
  \caption{Potential and Divergence in Road Network of Downtown LA at 8:30}
  \label{fig:LA 8:30}
\end{figure}

\begin{figure}[htbp]
  \centering
  \begin{subfigure}[b]{0.4\textwidth}
    \includegraphics[width=\textwidth]{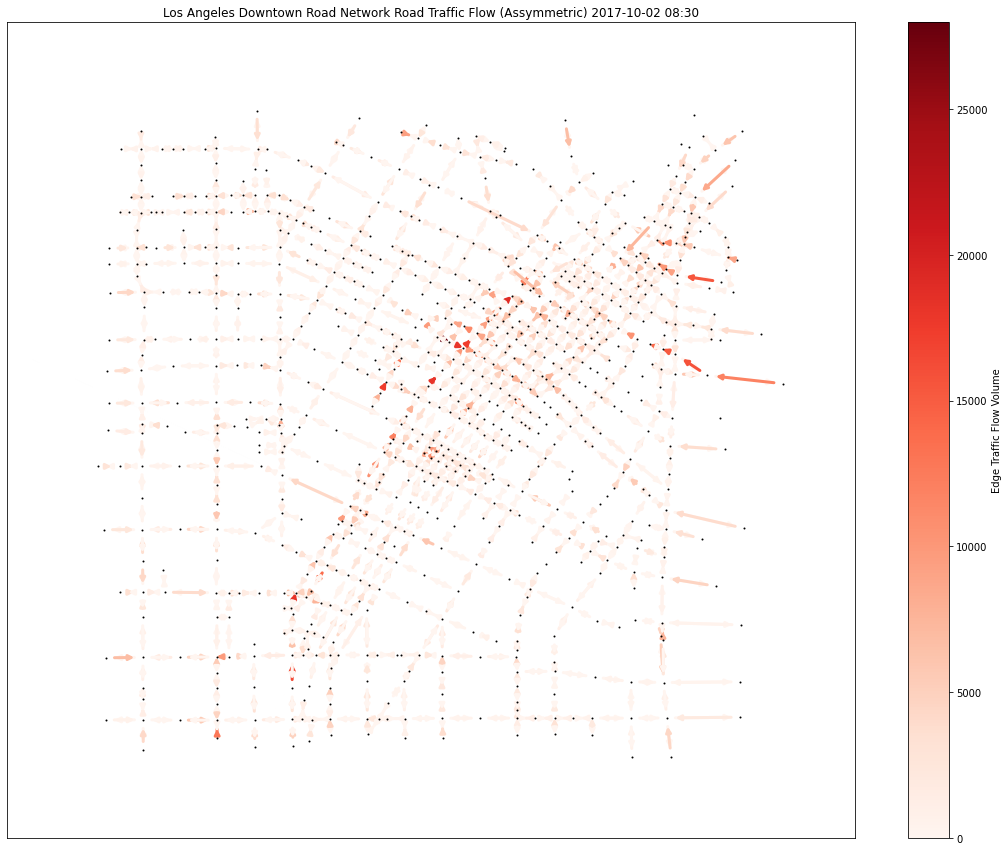}
    \caption{Edge Flow Raw Data in Road Network of Downtown LA at 8:30}
    \label{fig:LA 8 raw}
  \end{subfigure}
  \begin{subfigure}[b]{0.4\textwidth}
    \includegraphics[width=\textwidth]{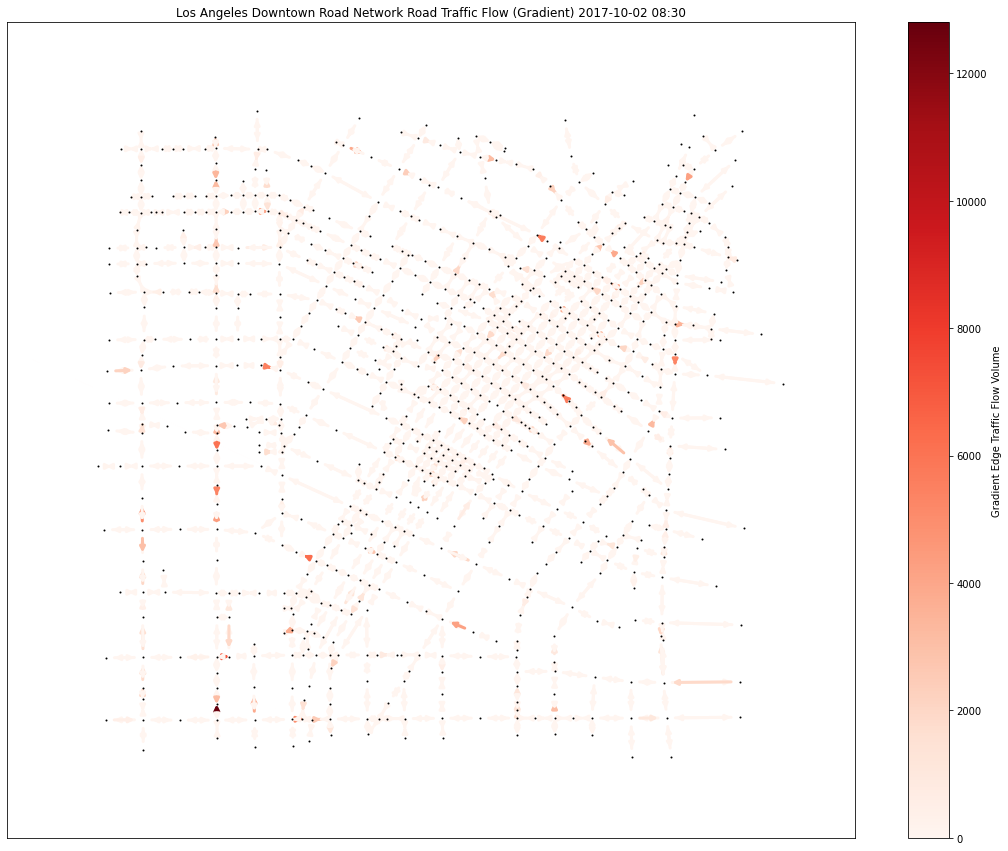}
    \caption{Gradient Component in Edge Flow in Road Network of Downtown LA at 8:30}
    \label{fig:LA 8 grad}
  \end{subfigure}
  \begin{subfigure}[b]{0.4\textwidth}
    \includegraphics[width=\textwidth]{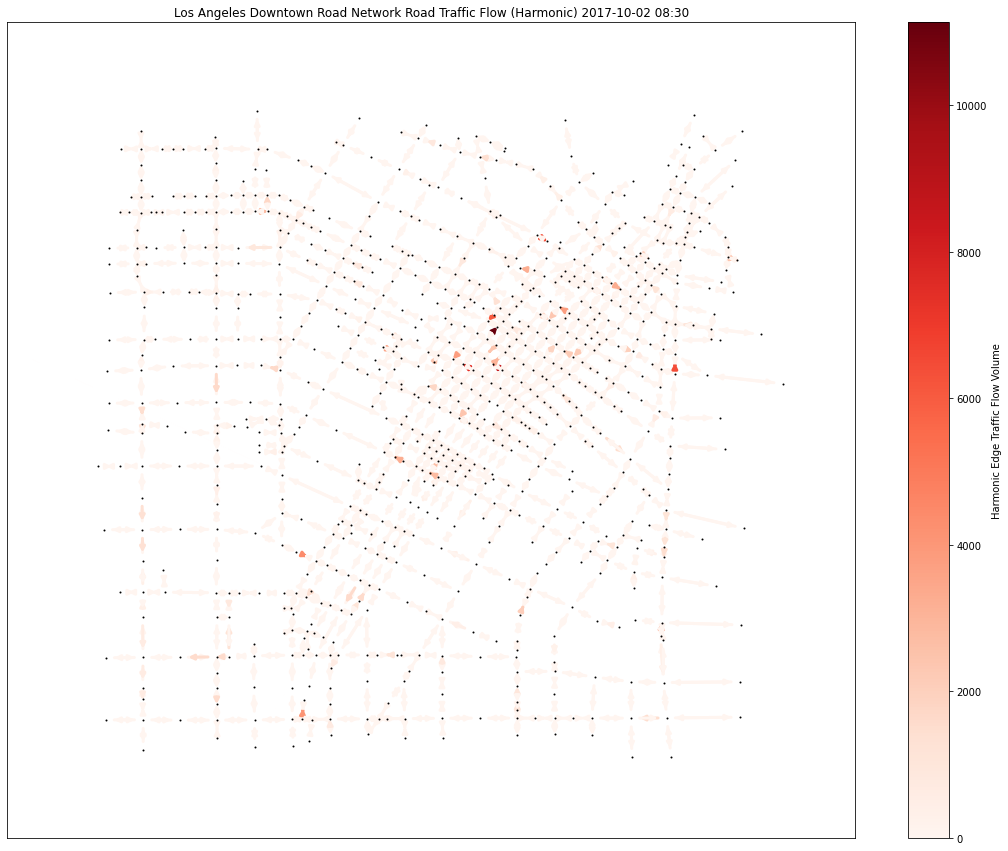}
    \caption{Harmonic Component in Edge Flow in Road Network of Downtown LA at 8:30}
    \label{fig:LA 8 har}
  \end{subfigure}
  \begin{subfigure}[b]{0.4\textwidth}
    \includegraphics[width=\textwidth]{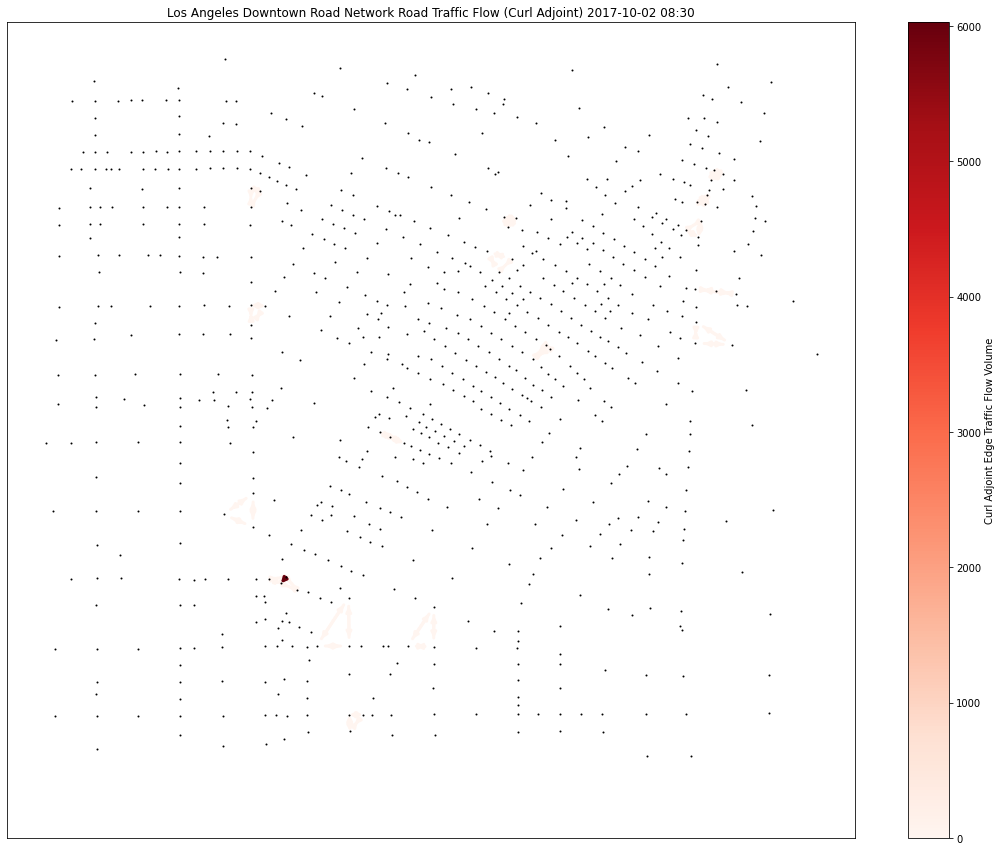}
    \caption{Curl Adjoint Component in Road Network of Downtown LA at 8:30}
    \label{fig:LA 8 curl}
  \end{subfigure}
      \caption{Hodge Decomposition in Road Network of Downtown LA at 8:30}
  \label{fig:LA hodge 8:30}
\end{figure}

\begin{figure}[htbp]
  \centering
  \begin{subfigure}[b]{0.4\textwidth}
    \includegraphics[width=\textwidth]{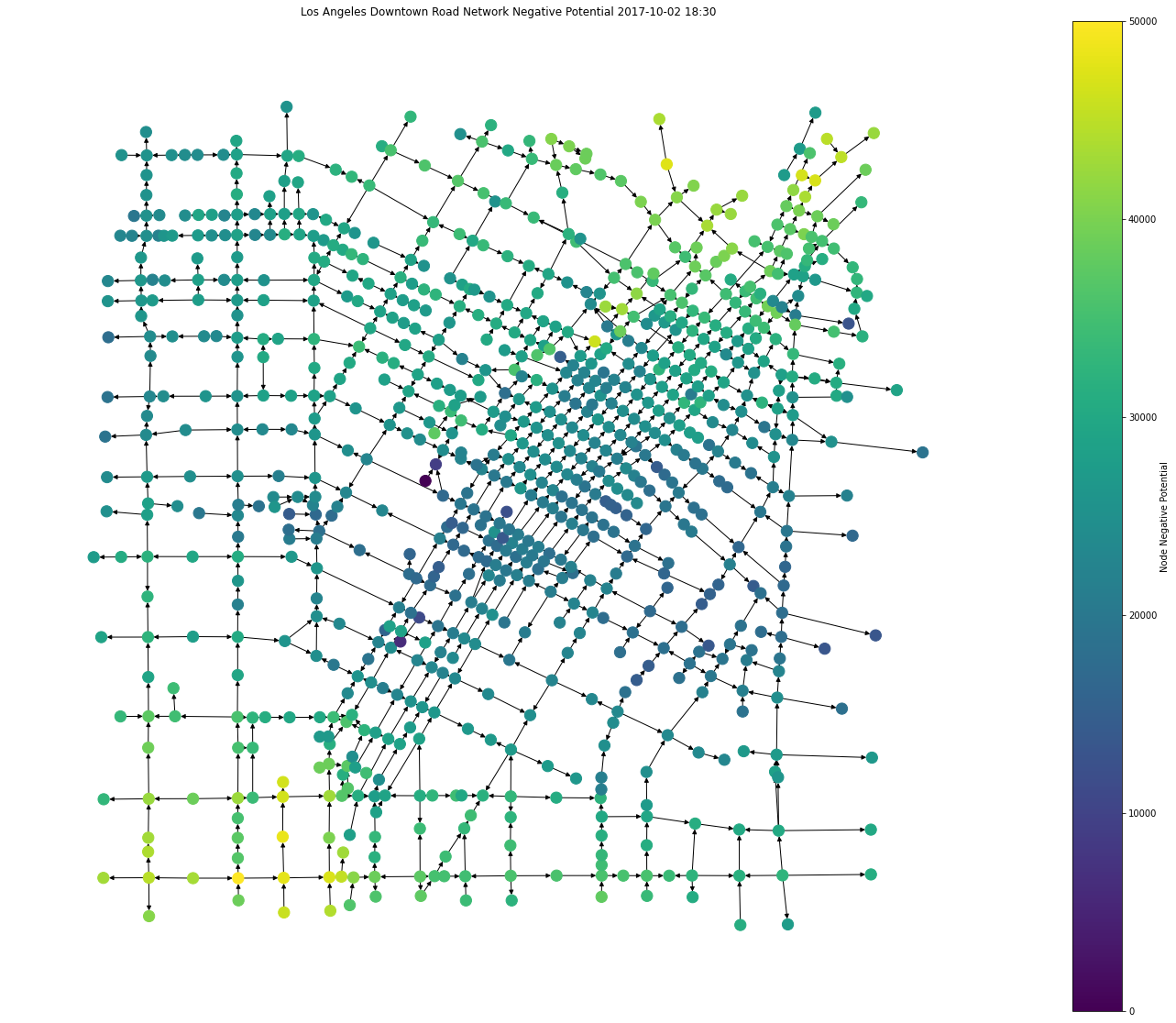}
    \caption{Potential in Road Network of Downtown LA at 18:30}
    \label{fig:LA 18 potential}
  \end{subfigure}
  \begin{subfigure}[b]{0.4\textwidth}
    \includegraphics[width=\textwidth]{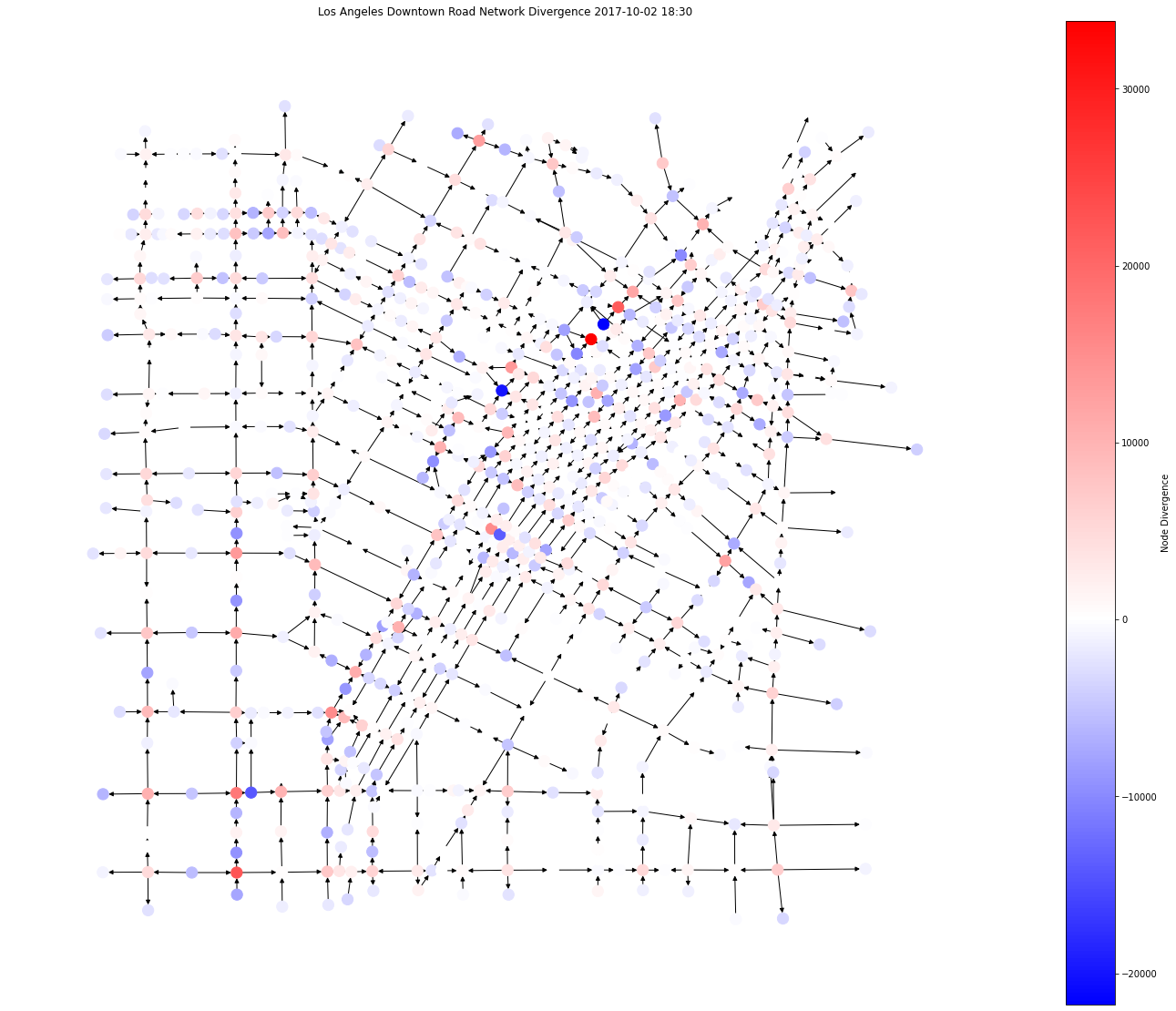}
    \caption{Divergence in Road Network of Downtown LA at 18:30}
    \label{fig:LA 18 divergence}
  \end{subfigure}
  \begin{subfigure}[b]{0.4\textwidth}
    \includegraphics[width=\textwidth]{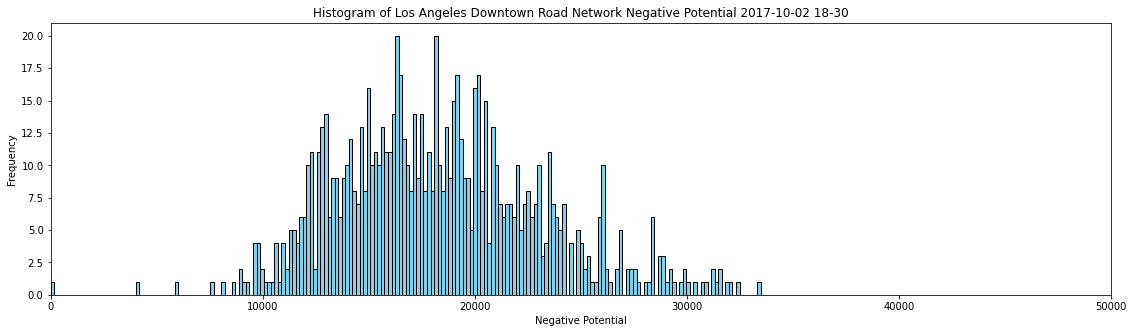}
    \caption{Potential Distribution in Road Network of Downtown LA at 18:30}
    \label{fig:LA 18 pot dist}
  \end{subfigure}
  \begin{subfigure}[b]{0.4\textwidth}
    \includegraphics[width=\textwidth]{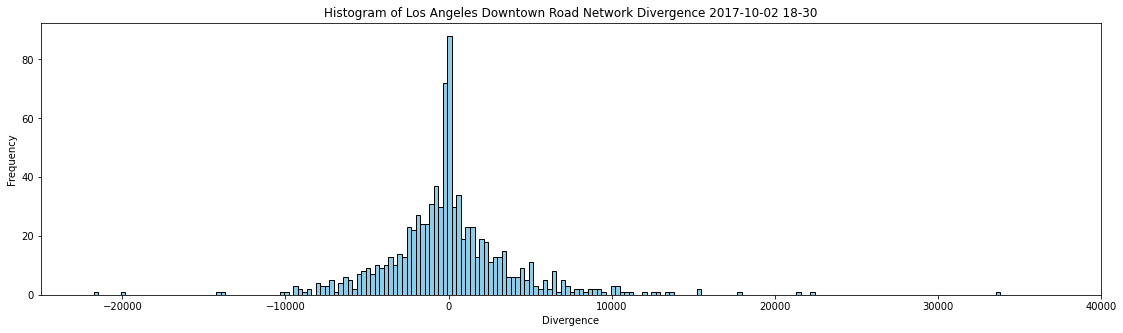}
    \caption{Divergence Distribution in Road Network of Downtown LA at 18:30}
    \label{fig:LA 18 div dist}
  \end{subfigure}
  \begin{subfigure}[b]{0.4\textwidth}
    \includegraphics[width=\textwidth]{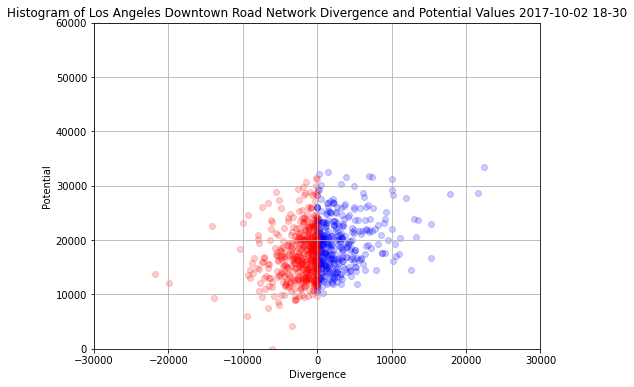}
    \caption{Potential and Divergence in Road Network of Downtown LA at 18:30}
    \label{fig:LA 18 pot div}
  \end{subfigure}
  \caption{Potential and Divergence in Road Network of Downtown LA at 18:30}
  \label{fig:LA 18:30}
\end{figure}

\begin{figure}[htbp]
  \centering
  \begin{subfigure}[b]{0.4\textwidth}
    \includegraphics[width=\textwidth]{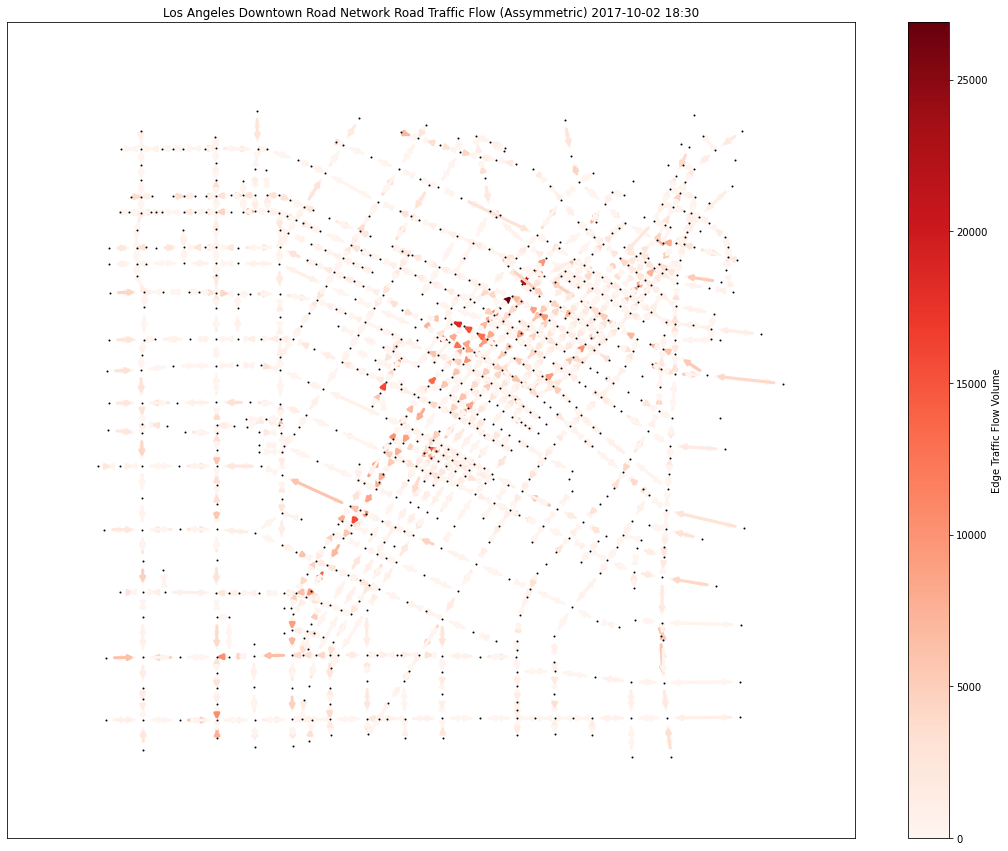}
    \caption{Edge Flow Raw Data in Road Network of Downtown LA at 18:30}
    \label{fig:LA 18 raw}
  \end{subfigure}
  \begin{subfigure}[b]{0.4\textwidth}
    \includegraphics[width=\textwidth]{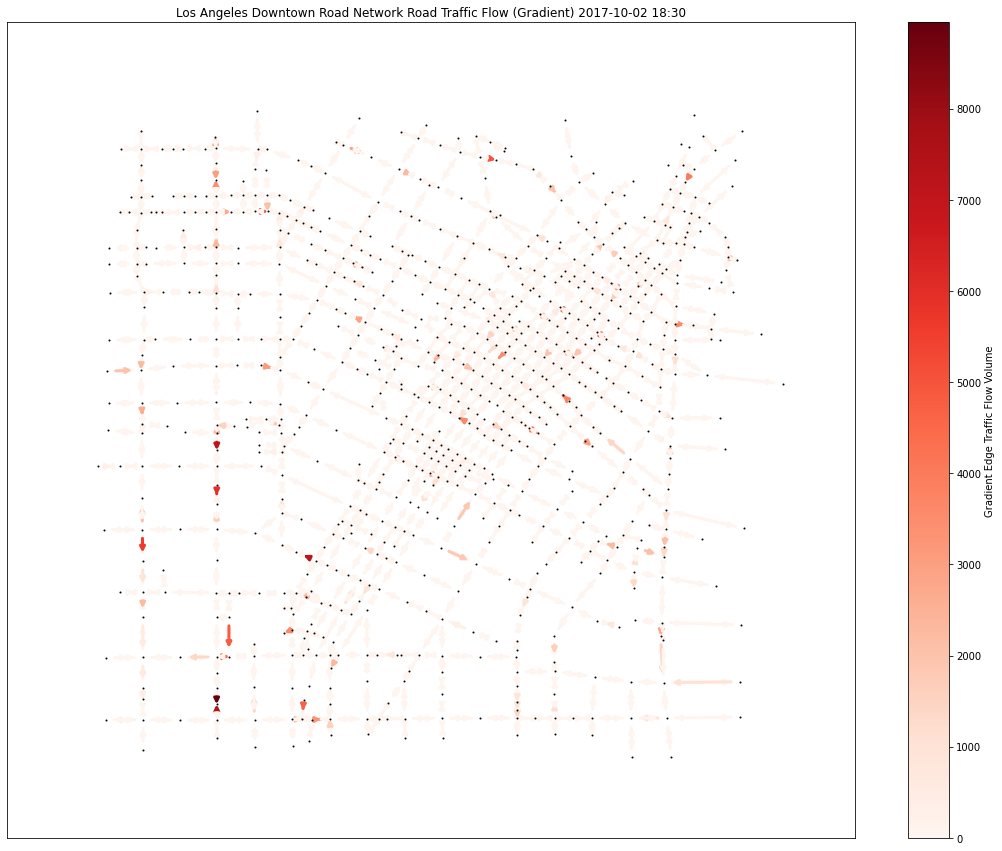}
    \caption{Gradient Component in Edge Flow in Road Network of Downtown LA at 18:30}
    \label{fig:LA 18 grad}
  \end{subfigure}
  \begin{subfigure}[b]{0.4\textwidth}
    \includegraphics[width=\textwidth]{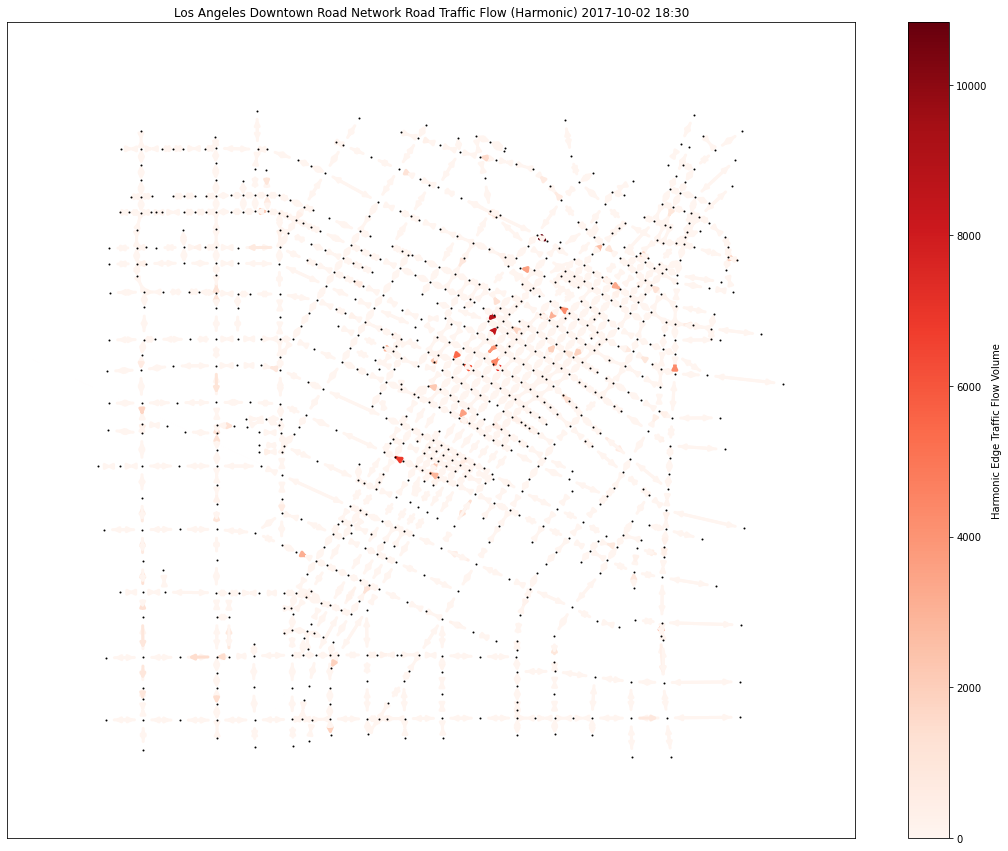}
    \caption{Harmonic Component in Edge Flow in Road Network of Downtown LA at 18:30}
    \label{fig:LA 18 har}
  \end{subfigure}
  \begin{subfigure}[b]{0.4\textwidth}
    \includegraphics[width=\textwidth]{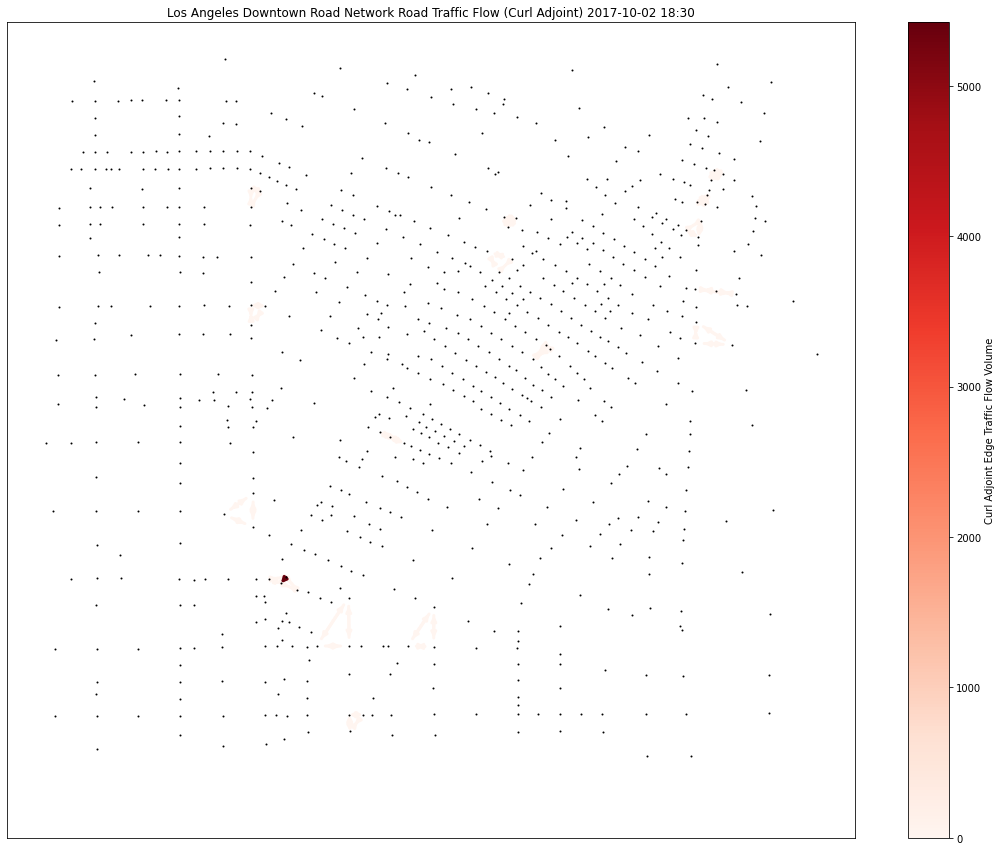}
    \caption{Curl Adjoint Component in Road Network of Downtown LA at 18:30}
    \label{fig:LA 18 curl}
  \end{subfigure}
      \caption{Hodge Decomposition in Road Network of Downtown LA at 18:30}
  \label{fig:LA hodge 18:30}
\end{figure}

\newpage

\section*{Appendix E}

\begin{figure}[htbp]
  \centering
  \begin{subfigure}[b]{0.4\textwidth}
    \includegraphics[width=\textwidth]{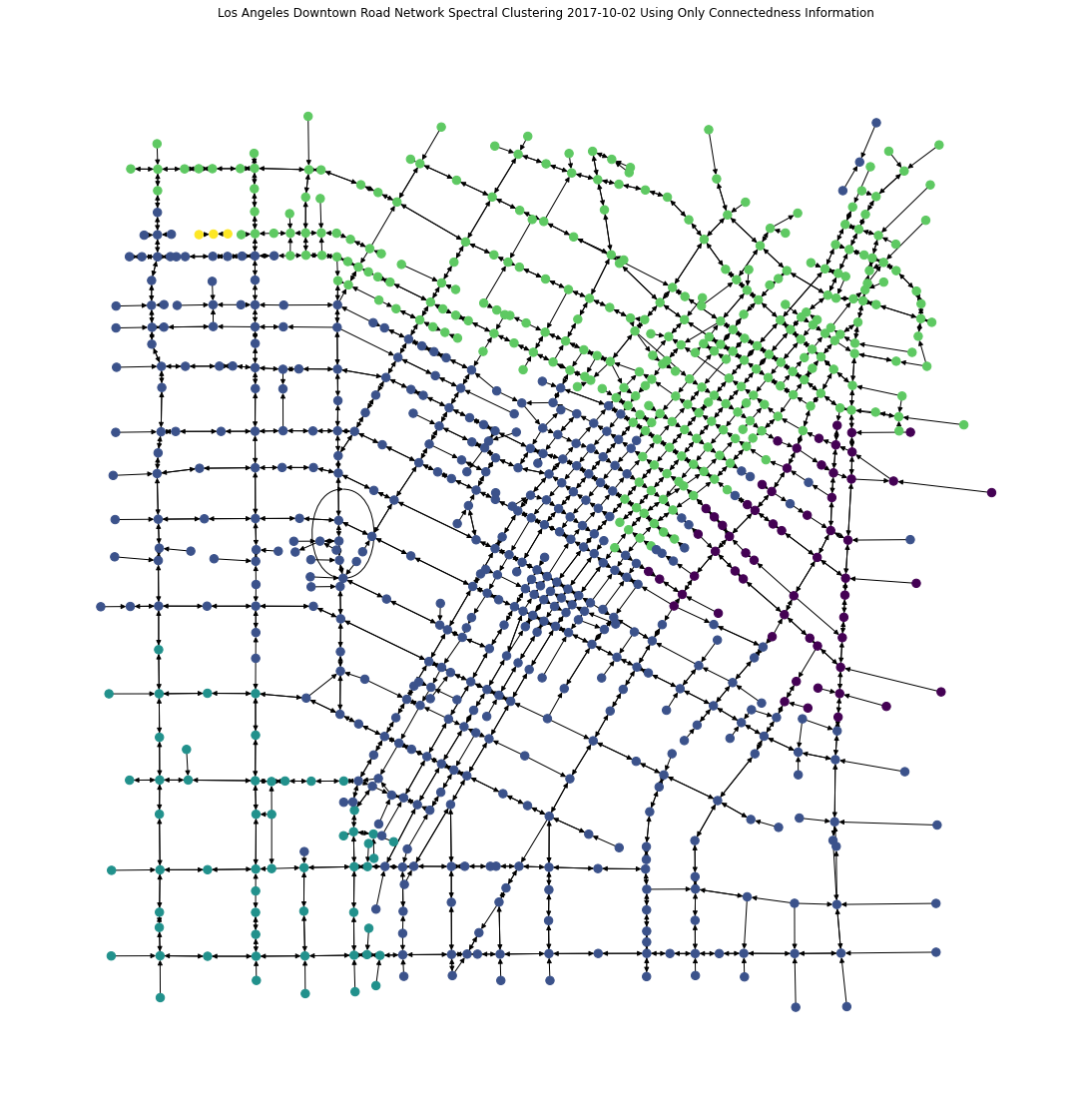}
    \caption{Clustering of Road Network of Downtown LA Using Only Network Structure}
    \label{fig:LA cluster 1}
  \end{subfigure}
  \begin{subfigure}[b]{0.4\textwidth}
    \includegraphics[width=\textwidth]{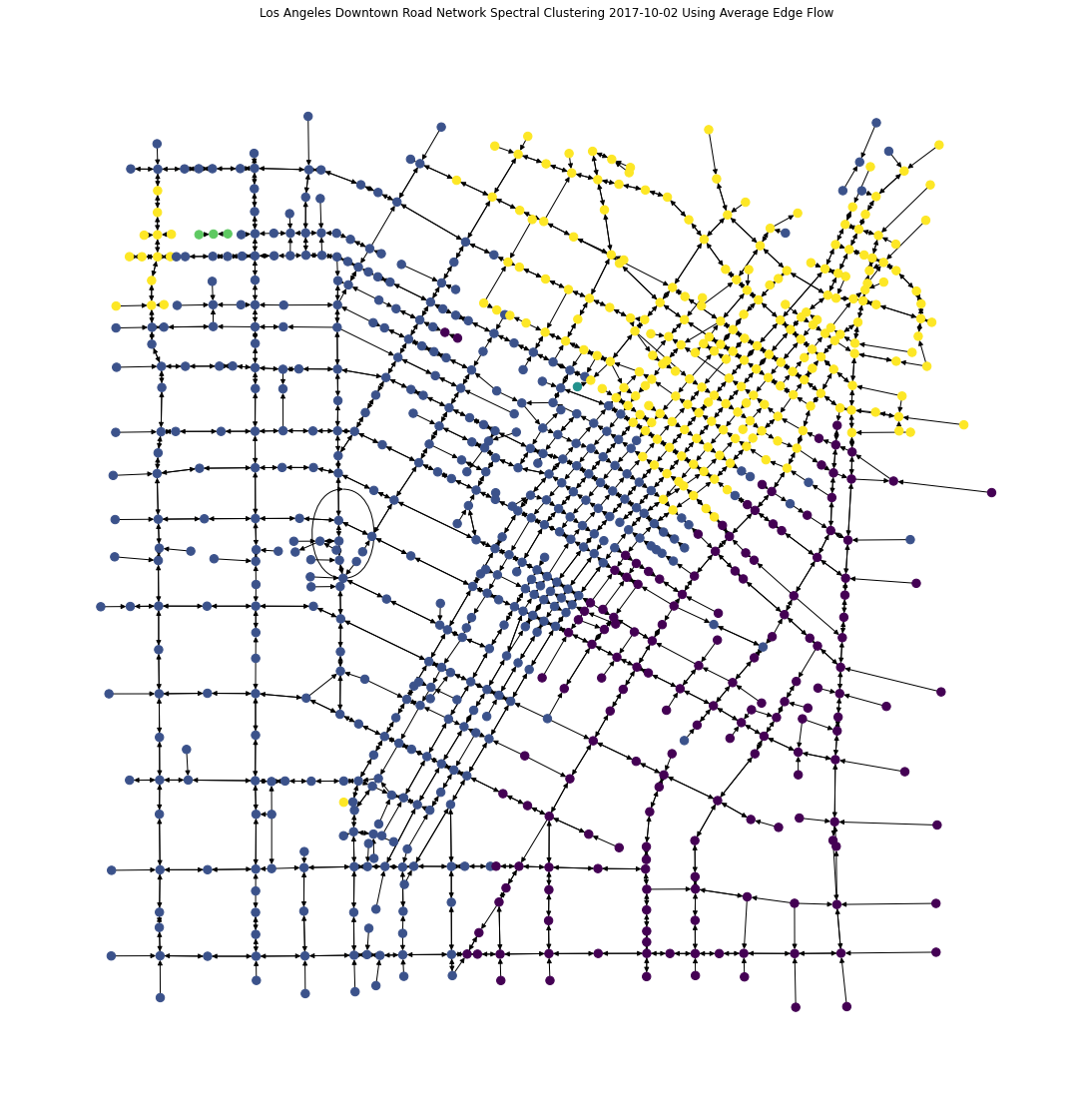}
    \caption{Clustering of Road Network of Downtown LA Using Only Mean Flow Volume on Edges}
    \label{fig:LA cluster 2}
  \end{subfigure}
  \begin{subfigure}[b]{0.4\textwidth}
    \includegraphics[width=\textwidth]{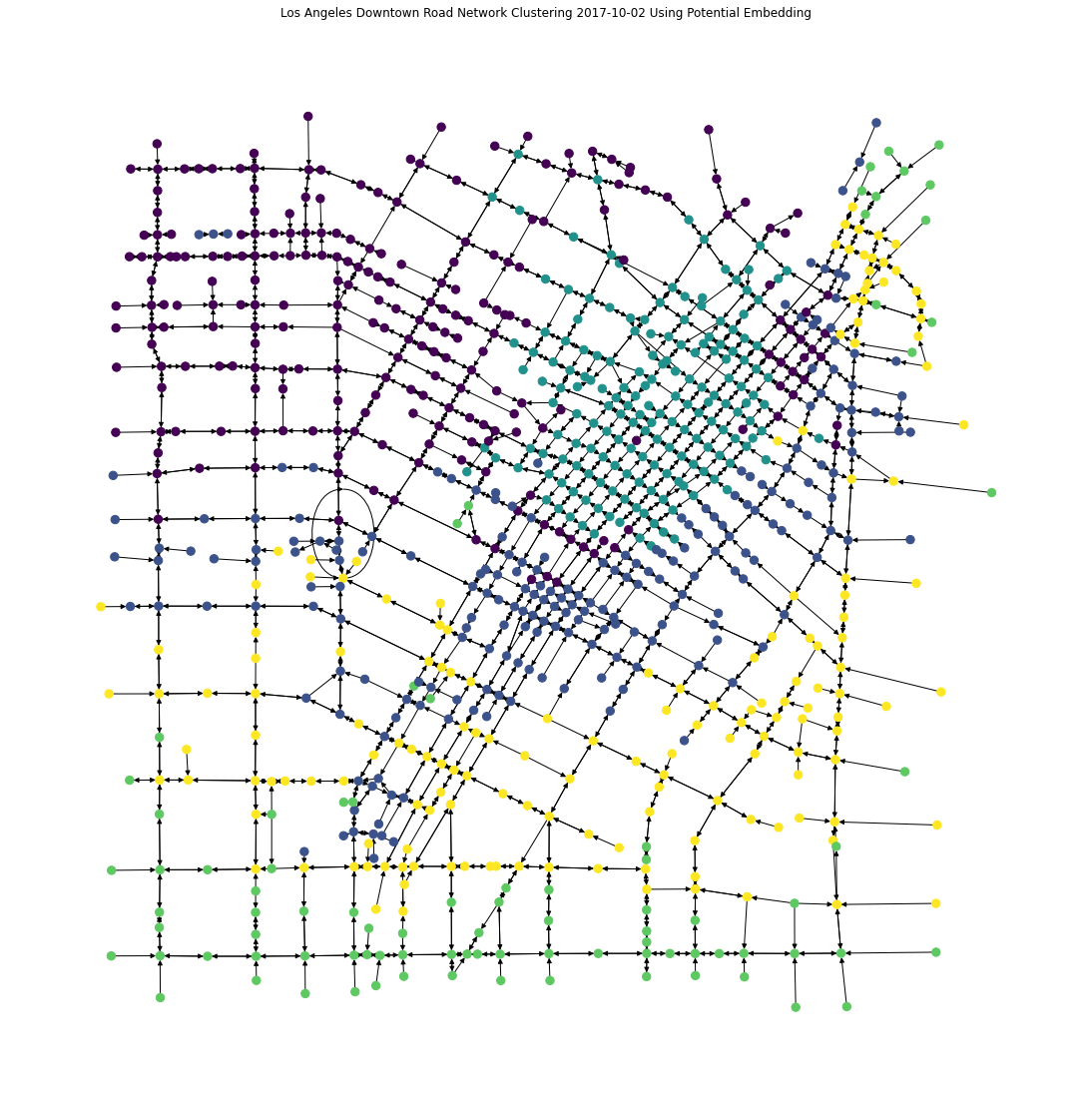}
    \caption{Clustering of Road Network of Downtown LA Using Only Potential Embedding}
    \label{fig:LA cluster 3}
  \end{subfigure}
  \begin{subfigure}[b]{0.4\textwidth}
    \includegraphics[width=\textwidth]{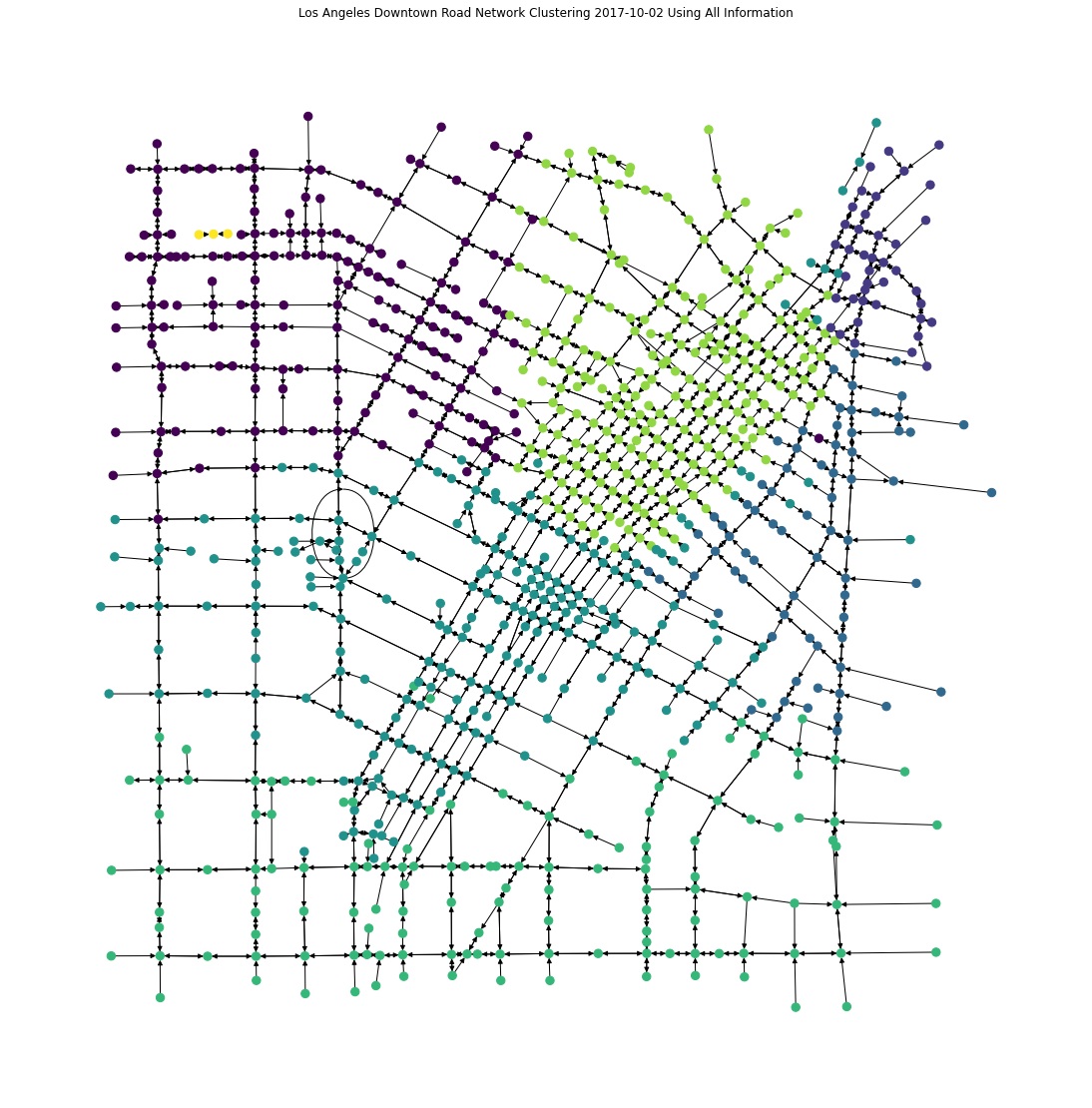}
    \caption{Clustering of Road Network of Downtown LA Using All Information}
    \label{fig:LA cluster 4}
  \end{subfigure}
      \caption{Clustering of Road Network of Downtown LA}
  \label{fig:LA cluster}
\end{figure}

\end{document}